# Quantifying the Vices and Virtues of Snakes and Ladders Through Time

Eleanor A. Dauenhauer[1], Paul J. Dauenhauer[2*]

[1] Chippewa Middle School, 5000 Hodgson Road Connection, St. Paul, MN, USA 55126
[2] University of Minnesota, Department of Chemical Engineering & Materials Science, 421 Washington Ave. SE, Minneapolis, MN, USA 55455
* Corresponding author: hauer@umn.edu

**Abstract.** The game of Gyan Chauper or 'snakes and ladders' exists in many forms throughout history as a board game of varying size, structure, and game elements of snakes and ladders associated with various vices and virtues inscribed within the board. Here, three boards were analyzed via Monte Carlo simulation, including the 1998 Milton Bradley version, the 72-square Vaisnava board, and the 84-square Jaina board, with the goal of understanding the relationships between board design and associated behaviors and spiritual concepts. Game play on each board was simulated 100,000 times with variations that included individual removal of a snake or ladder, thereby quantifying the importance of that element towards achieving victory. Comparison of the weighted importance of each game element and associated vice and virtue then permitted analysis of their importance for the game designer and their associated culture.

**Introduction.** Throughout history humans have engaged in games of skill and chance for entertainment, social development, and competition, providing parallel activity that sometimes mimicked or revealed the challenges of human life.[1] These games have taken many forms including physical competition, cards, ropes, dice, and surfaces including boards drawn in the earth, on fabric, or wood, all of which can represent aspects of human time including birth, growth, death, and beyond, with the rate and pathways of the represented life depicted in the structure of the game itself.[2] The distribution of game types and mechanisms of play are as diverse throughout the world as the different cultures that engage in them. The integration of these games in local cultures has taken on important roles in social integration, cultural diffusion, and even education, making them a key element of society and reflection of the behaviors and values of the players going back to ancient time.[3,4]

Games played on hard surfaces or 'board games' provide a unique record of historical cultures.[5] Played on a physical surface, aspects of the game are recorded into the board providing evidence for the rules of the game which reflect on that civilization. In ancient cultures that have buried their dead with treasures like jewelry, clothing, and game boards, these physical artifacts provide knowledge of the values, structure, and roles of that society. Many historical board games have represented the status of members of society and their mechanisms of conflict; this type of game R.C. Bell referred to as 'war games' and includes the chess group,[6,7] the Alquerque group,[8] and the related group of games called Draughts also referred to as 'checkers' developed around 1100 A.D.[5] In these games, the structure of the board, the positions and movements of the pieces, and the overall value and power of pieces towards victory reflect their importance in society and the meaning of play for the game players.[9,10,11,12]

The social meaning and potential educational value of board games has not been overlooked by the creators of games now or in history. A natural instinct for gameplay has been proposed as a core element in the development of society and human culture; as stated by Johan Huizinga, "For many years the conviction has grown upon me that civilization arises and unfolds in and as play."[13] More recently, Steven Johnson in his book, Wonderland, argued that play is also the origin of significant progress that has led to the modern world; the playing of games led to new ways of thinking and eventually new technologies, which ultimately changes society itself.[14] As a quote attributed to designer Charles Eames, "Toys and games are the prelude to serious ideas."[14,15] Play in the form of board games with their combination of spatial, mathematical, and social interaction has been shown to be an effective educational tool. For example, students developing basic numerical skills exhibit enhanced learning when playing linear number board games[16,17,18]. While the extent of learning occurring from the playing of board games is complicated, particularly of more complex social concepts such as behavioral norms, there is significant evidence that the content encoded within the game influences the players.[19]

Of interest here is influence of the Indian game of Gyan Chauper, more commonly known as 'Snakes and Ladders' or 'Chutes and Ladders.' This racing-style game consists of one of many possible boards with an array of numbered squares varying from ~60 to several hundred; players use dice, shells, spinners, or other random methods to initiate movement along the squares allowing them to move from the bottom initial square to the final top square and victory.[20,21] Progression is manipulated through mechanisms that accelerate or slow players towards the ultimate final square. After moving squares, players that land on the head of snakes or the top of chutes (i.e., slides) move downward to lower squares. Alternatively, players at the bottom of a ladder can move upward and speed their progress toward victory.

While the steady progression through the game board associated with random dice rolls (or spins or shells) represents growth through life, the interesting elements of Gyan Chauper are the snakes/chutes and ladders that take on meaning with the descriptive labels and design of the game board. It has been observed that strategy games frequently relate to social systems, while games associated with physical skill and chances are associated with environmental conditions and religious beliefs.[22] Gyan Chauper follows this pattern; player movement and chance at victory over other players progressing through the game is completely due to chance, similar to many other historical games that are spiritual in nature.[23] As described by Topsfield, the game's early theme was "the spiritual quest for liberation from the vicissitudes of karma or the hindrances of the passions."[20] Climbing a ladder upward is associated with good actions (virtue) or positive karma, while a snake will pull the player downward due to bad actions (vices) or negative karma. In some versions of the game, victory is achieved by reaching the top of the board known as Vaikuntha (the heaven of Visnu), and players can be pulled down into the board reliving parts of their lives (traversing squares on the board) multiple times before reaching the afterlife.

The actions of vice or virtue and karma can be found described on the boards of Gyan Chauper through history, as carefully documented via preserved game boards, photographs and drawings of historical boards, and descriptions of game structure and play.[24,25] The game that was invented in India has changed through time with varying numbers of squares, structure of the squares, the length and distribution of snakes and ladders, and their associated vices and virtues.[26] Three games are presented in **Scheme 1A-1C** including the 1998 Milton Bradley Chutes & Ladders 100-square game board, the 72-square Vaisnava board, and the 84-square Jaina board, respectively. While the modern board is publicly available, the history and gameplay rules of the older 72- and 84-square gameboards have been extensively examined in the thesis of Schmidt-Madsen, where he provides a history and cultural influence.[27] As he describes the content of the game, the vices and virtues of the older 72-square board relate behaviors such as hatred or compassion to penalties or rewards such as greed or the Realm of Brahma, respectively, via the connections on the board of snakes or ladders.[27] Similarly, the 84-square Jaina board connects virtues and vices like meditation and envy with rewards and penalties such as omniscience and jealousy.[27] The modern Milton Bradley board has significantly altered the tone of the game changing it from a progression throughout life to a shorter period of childhood, with the vices and virtues being similarly reduced to stealing cookies or rescuing a cat, ultimately removing all spiritual elements of the game in favor of the cause-and-effect relationships of activities in childhood.

By integrating behaviors and associated penalties and rewards into the game boards, the magnitude of their role towards victory can be quantified via simulation. For each behavior, the penalty or reward presumably corresponds to the movement of the player by a specific number of squares (forward for rewards, backward for penalties). However, the placement of snakes and ladders in addition to their interaction with each other significantly alter the extent of the penalty or reward, such that the actual benefit of any snake or ladder can only be statistically assessed via playing the game. In this work, the game of Snakes and Ladders is simulated for each of the three boards of **Scheme 1** via the Monte Carlo method, and the average number of spins or dice/shell throws required to achieve victory is assessed for 100,000 trials of the game. The magnitude of each snake or ladder is then determined via comparative simulation of the

_________________________________________________________________________________

game via systematic removal of each individual snake or ladder, ultimately determining how many additional or fewer spins or dice/shell throws are required to achieve victory. Games are simulated by the rules and boards outlined by Schmidt-Madsen,[27] permitting a quantitative analysis of the importance of different behaviors encoded in the boards throughout history.

**Methods.** Three game boards were simulated via the Monte Carlo method using the game boards described by Schmidt-Madsen and depicted in Scheme 1. Simulation codes are provided in the supporting information.

*1998 Milton Bradley Chutes and Ladders Game Board.* The board of gameplay consists of 100 ten-by-ten squares. Chutes and ladders are distributed across the board, and their placement has varied throughout different versions of the game with time. The square 1 that your player begins on is located in the lower left corner of the board, and the square 100 that you end the game on is placed in the upper left corner. Each square is numbered through 1 to 100 to track player progress from the beginning to the 100th square. You can view the placement of the ladders and the chutes in **Figure 1A**. This version of the game generates random movement via a spinner, which consists of the shape of a circle with six equal parts labeled 1-6 and has an arrow that can rotate between all six equal parts.

To begin a game of Chutes and Ladders, a character or token is first placed to the left of the first square on the board ('square zero'). The objective of the game is to reach square 100, which can be achieved by steadily progressing your player forward with each spin or by landing on the ladder that starts at space 80. Players advance in each turn by the number of squares identified by the spinner, but spins that would result in players moving past square 100 result in no movement. Suppose you reach space 10, and you have spun a 3. You would no longer move to the right, given that in doing so you would move off the board. Instead, move your character up one, to space 11, and then to the left two, so as to reach 13. You repeat this process for the rest of the game, alternating left or right depending on which side of the board you are currently at. The ladders of the game take you higher on the board, and the chutes decrease your position. Say you are at space zero, and you spin a four. This would land you on the cake-baking space, and you would follow the ladder to wherever it would lead you on the board, in this case, moving to space 14. In the case of landing on the top of a chute, you would follow it down the game board and keep your player at the space that it ends on.

*72 Square Snakes and Ladders Game Board.* The 72-square board game begins with each player having their token on square zero (off the board). The players progress along the board by a specific number of squares indicating by change. It should be noted that the description by Schmidt-Madsen indicates that players would use 7 Cowrie shells, allowing for moves of 0-7 squares. We have simulated the game with seven Cowrie shells assuming that each shell has equal probability of landing up or down. We have also simulated games with a six-sided die for better comparison with the 100-square board game. Both sets of game simulations, with a six-sided die or with 7 Cowrie shells, are presented side by side. Players proceed up the game board with each throw of the die/shells; landing on the base of a ladder or the head of a snake after moving results in players being moved up or down the board, respectively, according to the length of the particular ladder or snake. Victory is awarded to the player that first lands on square 68, which can be reached by continuing to increase in squares with each throw of the die/shells or by landing on the ladder at square 54. For players that overshoot square 68, they must then land on the head of the snake at square 72 and proceed to square 51, after which on subsequent turns they can again attempt to land on square 68.

*84 Square Snakes and Ladders Game Board.* The 84-square board game begins with each player on square 1. Players throw a stick die of four sides with options to move 1, 2, 5, or 6 spaces. Players can only proceed beyond square 1 after throwing a 1 on the stick die which brings them to square 2. Once they arrive at square 2, they continue to advance through the game boards with each throw of the stick die. During the game, if a player lands at the head of the snake, then they immediately proceed to the end of the snake during that turn. Alternatively, if a player lands at the base of the ladder, they do not proceed; on the next turn, they move up the ladder if they roll a 1 on the stick die or move beyond the ladder with a roll of 2, 5, or 6. Victory is achieved by reaching top square #6, located at the top of the game board. The top six squares are only accessible via the ladders at squares 47, 50, or 80. In the course of the game, if players overshoot all possible ladders that would allow them to ascend to the top squares, then they proceed until reaching the snake head at square 76, which then brings them down to lower squares and allows them to potentially land on a ladder again which will bring them up into the top square region. Once players have exceeded square 80, there are no possible routes to ascend to the top squares; for this reason, players that reach square 84 must return backwards counting down from 83, 82, 81…until reaching the snake at square 76 or by landing on the base of the ladder on square 80. Once players have ascended into the top square region, they advance only with the roll of a 1.

*Simulation Methods.* Each game board was simulated in Python within Microsoft Visual Studio Code version 1.76.1. Simulations accounted for the progression of a player's token beginning from the starting position (0 or 1) and progressing up through the board with each random spin or shake value, generated randomly in each turn. Simulations were conducted with only one person playing the game. For each turn, the implementation of a chute/snake or ladder was assessed either at the end of the turn or at the start of the turn, depending on the considered board game variation. The number of spins or shakes of dice/shells required to complete each game was counted and recorded. Each game condition was simulated 100,000 times, and the data were saved and loaded into Microsoft Excel where it was analyzed for statistical variation.

*Statistical Analysis.* The variations of the game boards was assessed by simulating the game 100,000 times for each possible game board arrangement. For each simulation set, the number of spins or die/stick throws required to win the game was recorded and stored 100,000 times. This data set was then evaluated to assess the average, median, minimum, and maximum. Data was also sorted as a histogram to depict the distribution of the number of spins required to achieve victory. Tables of summary data are provided in the supporting information file.

**Results and Discussion**. The three game boards of **Scheme 1** were evaluated by a Monte Carlo simulation of 100,000 games for each game condition with only one player ascending the board to reach the victory square (e.g., 100). The initial simulations evaluated the game in its original form, with the removal of all snakes/chutes, the removal of all ladders, and a fourth scenario with all snakes/chutes and ladders removed. This permitted an overall evaluation of the importance of snakes/chutes and ladders towards the speed of achieving victory. Thereafter, a set of scenarios were considered which removed individual snakes/chutes or ladders for each set of 100,000 simulations, thereby permitting a quantitative analysis of the importance of each individual game element.

*Milton Bradley 1998 Game Board.* The 1998 Chutes and Ladders Milton Bradley game board is depicted in a photograph in **Scheme 1A** and in a diagram in **Figure 1**. Comprised of a 100-square board, players start off the board (square zero) and progress upward to victory at square 100 via movements of 1-6 squares determined from a spinner. In the game board of **Scheme 1A**, all of the ladders are associated with virtuous behavior such as planting flowers, cleaning up, rescuing a cat, or returning a lost purse. Alternatively, all of the chutes are associated with naughty behavior such as pulling a cat's tail, riding a bicycle without hands, skating on thin ice, or reading comic books. All chutes and ladders are listed in **Table 1** along with their associated vice/virtue, the square numbers that they connect, and an assigned letter A-S to refer to each game element. The letter assignments of each chute and ladder are depicted in **Figures 1B** and **1C**, respectively.

The simulation of 100,000 games of the Milton Bradley 1998 game resulted in the histogram of **Figure 2A** depicting the frequency of games requiring a total number of spins for victory between 1 and 300. The game can be won in as few as 7 spins, which only happens in <0.3% of games. The most common number of spins to win the game was 19-20, which happens in about 5% of games. However, the histogram has a longer tail at higher numbers of spins making the average of ~40 spins (and median of 33 spins). About 96% of games are completed within 100 spins.

The duration of the game changes considerably when removing chutes or ladders as shown in **Figure 2B-2D**. The removal of all chutes and ladders from the game board yields a histogram with a narrow number of spins; the most common number of spins to win the game is 31-32 which occurs in ~17% of games. About 98% of games are completed within 50 spins. Alternatively, the game with chutes but without any

---

ladders exhibited a large distribution of spins required to win the game; the average and median number of spins was 120 and 93, respectively. In contrast, the game with ladders but no chutes was fast with an average and median number of spins to win the game of 22 and 21 spins, respectively.

For the 1998 Milton Bradley game board, the contribution of individual game elements was examined by two different methods. In the first set of simulations, individual chutes (**Figure 3A**) or ladders (**Figure 3B**) were added to a completely blank 100-square board, and each scenario of the game with a single chute or board was simulated 100,000 times. The addition of chutes to a blank board lengthened the time required to play the game; as shown in **Figure 3A**, the average number of added spins required to win increased from ~0.25 to ~7 added spins relative to a completely blank board. There existed a linear relationship between the length of the chute (measured in tiles) and the average number of added spins, with only chute S which moves players from tile 98 to 78 (for pulling a cat's tail) exhibiting a penalty disproportionate to its chute length. When individual ladders were added to a blank 100-square game board (**Figure 3B**), again there was correlation such that the longest ladders reduced the average number of spins per game by up to 5 spins. Ladders A and I both deviated from the general trend observed with all ladders. Ladder A associated with planting a garden only reduced the game length by 1.75 spins on average despite being a 37-tile ladder; in contrast, ladder I associated with winning a pet show reduced the game by 2.75 spins on average despite only being 20-tiles in length.

The interaction of ladders and chutes in gameplay means that individual game elements cannot be evaluated independently as depicted in **Figure 3A** and **3B**. Instead, we will also assess the contributions of individual chutes or ladders by removing them from the full game board; each set of 100,000 simulations will then compare the full game board simulation with the game missing a single game element (ladder or chute). As shown in **Figure 3C**, removal of individual chutes from the full game shortens the game by reducing the average number of spins by ~7-8 spins. Now most chutes are close to a linear correlation between chute length and average added spins except for chute N (associated with breaking dishes), which has a disproportionately low impact on gameplay (-3 average spins) despite being a chute of length 43 tiles. Similarly, individual ladders were removed from the full game board and simulated as shown in **Figure 3D**. By this method of comparison, removal of ladder I adds ~19 spins to the game, which is almost double the next most important ladder E (~10 added spins).

*72-Square Vaisnava Board.* The Vaisnava board depicted in **Figure 4A** is comprised of 72 tiles/squares, 10 ladders, and 10 snakes. Snakes are labeled in **Figure 4B**, and ladders are labeled in **Figure 4C**. All snakes and ladders and their associated vices, virtues, rewards, and penalties are listed in **Table 2**. Starting off the board, players throw seven Cowrie shells to advance along the board until they reach the square of 68 to win the game. Unique to this board is the placement of the winning 68-numbered square, which is in the middle of the top row of the board. Players can win the game by landing on square 68, but if they overshoot then they need to proceed to square 72, where snake T will move them back down to square 51.

Simulation of the 72-square Vaisnava board of **Figure 4A** was conducted for the full game with all snakes and ladders with 100,000 trials of four scenarios depicted in **Figure 5**. All simulated conditions were considered with either a six-sided die (**Figure 5A-5D**) or 7 Cowrie shells (**Figure 5E-5H**) to evaluate the differences resulting from the varying probability of token movement values. The histogram of the distribution of throws of a six-sided die (referred to as "spins" from here on) for the full game in **Figure 5A** indicates that the minimum number of spins is 5, which only occurs in ~0.5% of all games. The most common number of spins is 11-12, which occurs in 3.3% of games. Thereafter, a long tail exists with only about 80% of games being completed in less than 100 spins. The elimination of all ladders and snakes from the six-sided die simulation (except snake T required for the game to proceed) in **Figure 5B** significantly changes the histogram of spins. A significant number of games (43%) are won with about 30 spins or less, after which a smaller tail of spins occur associated with the player cycling through the upper game via Snake T. When all ladders are removed (**Figure 5C**), the game played with a six-sided die can only be won in 13 spins, after which there is a broad distribution of spins extending out far beyond 300 spins per game. Finally, removal of all snakes except T while retaining all ladders yields a narrow histogram (**Figure 5D**)

_______________________________________________________________________________

with ~96% of all games being completed with <100 spins. Statistics of all four scenarios are provided in the supporting information.

Parallel simulations of the 72-square Vaisnava board were conducted using 7 Cowrie shells with the results in **Figure 5E-5H**. By this method of selecting the movement of a token, each Cowrie shell was treated as an independent event; all shells could land up (+7 tile movement), all shells could land down (0 tile movement), or shells could land in any combination of up and down in between (+1 to +6 tile movement). Unlike the equal probability of movements that exists with a six-sided die (each of the six die sides are equally probably), seven independent Cowrie shells will exhibit a binomial distribution of possible movements. All shells landing up or all shells landing down corresponding to movements of +7 or 0, respectively, will be rare, while movements that can occur via multiple combinations of shell outcomes (e.g., 3 or 4 up shells) will be much more common. This distribution of board token movements will ultimately be determined by the probability, *P*, of any one shell landing up or down, which will depend on the specific geometry shells selected for play. To determine this, experimental data previously collected via the throwing of four shells was fit to a binomial probability distribution with variable, *P*, changed until the fit to the data was optimal (i.e., minimal model-data error); additional details provided in the supporting information.[28,29] The outcome of fitting this experimental data was that P was 50%; experimental evidence indicated that shells were equally likely to land up or down in any throw. For the purposes of the simulation of the 72-square Vaisnava board, movements were then determined with the following tile progression probabilities as described in the supporting information: 0 (0.8%), 1 (5.5%), 2 (16.4%), 3 (27.3%), 4 (27.3%), 5 (16.4%), 6 (5.5%), and 7 (0.8%).

The simulation of four conditions of the 72-square Vaisnava board with 7 Cowrie shells (**Figure 5E-5H**) yielded similar results to the simulations using a six-sided die (**Figure 5A-5D**), albeit with more 'spins' (tossing of the shells) per game to achieve victory. As shown in **Figure 5E**, a large number of games with the full 72-square Vaisnava board occurred in less than 200 spins, such that the average number of spins was ~100. However, the large tail of games that took longer than 200 spins increased the value of the average relative to the median number of spins of only 69. Without ladders or snakes (except the required snake T), the game is significantly shortened to an average of only 59 spins (median of 39 spins); as shown in **Figure 5F**, a large number of games are completed in fewer than 26 spins. When the board only contains snakes without any ladders (**Figure 5G**), the distribution of game spins is quite large with an average of 333 spins and some games requiring over a thousand spins. Finally, boards with only ladders and snake T are fast with an average number of spins of only ~53 as shown in **Figure 5H**. When compared side by side, games played with Cowrie shells or six-sided die are similar in their gameplay.

The contributions of snakes towards victory in the 72-square Vaisnava board using a six-sided die was assessed by removing individual snakes and simulating the game for comparison of the entire game with all elements as shown in **Figure 6A**; the alternative evaluation of individual addition of game elements was not considered due to its limited provided insight. All snakes except for snake N slow down a player's progress towards victory; despite being 17 tiles long, presence of snake N (associated with injury leading to hell) actually speeds up play by reducing the number of spins required to win the game by 0.3 spins. The other unique feature of snakes in the 72-square Vaisnava board is snake K (associated with darkness and anger), which is 60 tiles long and reduces the number of spins required to win by ~11 once it is removed from the board. This penalty for snake K is disproportionately more severe relative to its length when compared with all other snakes in the game.

The impact of ladders in the 72-square Vaisnava board using a six-sided die is complex with an inverse relationship between ladder length and impact on game speed as shown in **Figure 6B**. For most of the ladders of length 10-30 tiles, removal from the full game adds 1-6 spins on average. However, ladder J, one of the shortest ladders in the game only 14 tiles long associated with devotion and ascendance to Visnu's Heaven, has a significant impact on the game. When ladder J is removed from the full game, the number of average spins required to win the game increase to ~18 spins. Not surprisingly, ladder J directly moves players to the winning square 68. Similarly, ladder B associated with compassion and ascendance to the Realm of Brahma has a counterintuitive role in the game; despite being the longest ladder at 52 tiles in length actually speeds up the game by ~3 spins when it is removed from the game. This ladder, though a

___

considerable benefit to players in terms of overall increase in tile position, moves players to square 69 which is after the winning square of 68; players that move up ladder B are automatically required to proceed forward to slide 72, down snake T to tile 51, and then back up to the top of the board to attempt to win the game again.

The role of individual snakes and ladders in the 72-square Vaisnava board was also assessed using 7 Cowrie shells as shown in **Figure 6C-6D**. Individual snakes were removed to determine the average number of added spins relative to the length of the snake; as expected, the relationship was proportional as removal of longer snakes yielded a significant reduction in the number of spins required to achieve victory. When individual ladders were removed (**Figure 6D**), the 72-squre Vaisnava board played with Cowrie shells again exhibited complex behavior nearly identical to the game played with a six-sided die. Games played with Cowrie shells or six-sided dies were compared in **Figure 6E-6F**. The four considered scenarios exhibited comparable behavior for both the mean and median number of spins required to achieve victory, independent of the use of Cowrie shells or a six-sided die. Similarly, in the parity plot between the removal of snakes and ladders with either Cowrie shells or six-sided die, the impact on the number of pins added to the game to achieve victory yielded a linear relationship; the only impact on switching between movement mechanisms was a lengthening of the overall game when using Cowrie shells.

*84-Square Snakes and Ladders*. The 84-square snakes and ladders board depicted in **Figure 7A** can be interpreted as three games in one. Starting in square 1, players must first roll a 1 on a stick die with only four options to move (+1, +2, +5, +6); this will be referred to here as a "spin." Once the player advances beyond square one, they will continue rolling the stick die to move up the board. As they progress, they can end up at the end of a turn on the bottom of a ladder, as shown in **Figure 7B**; at the start of their next turn, they will advance up the ladder only if they 'spin' a 1. Alternatively, players can end their turn at the head of a snake, which immediately pulls them down to the tile with the tail of the snake, as shown in **Figure 7C**. The first two parts of the game, square 1 and progress up the board, continue until the player lands on ladder I, J, or O, which moves them up into the top region of the board. Players win by reaching top square 6 either directly from ladder I, or they can win by proceeding to top square 1 and moving up a square every time they spin a +1 and advance one tile. The 84-square snakes and ladders game has an additional complexity; the 84$^{th}$ game tile is in the top right of the board. Once a player is in the top row, they can continue moving with each spin, but if they overshoot tile 80 then they will continue on to tile 84. In this gameboard design, the player will then continue backwards (84, 83, 82, …) until they either land on tile 80 (and proceed up to top square 1) or until they get back to square 76 and proceed down snake X.

The 84-square snakes and ladders game board was simulated as shown in **Figure 8**. The full game with all snakes and ladders was simulated 100,000 times and the histogram of spins required to win the game by reaching top square 6 is depicted in **Figure 8A**. While the full game required an average and median number of spins of 110 and 82, the shape of the histogram with a long tail at higher spin games indicated that the most frequent game only had 31-32 spins. Games requiring fewer than 200 spins comprised 86% of all games. In **Figure 8B**, the game was again simulated 100,000 times with the removal of all snakes except X and all ladders except O; these two game elements were required to allow the game to continue to completion. The histogram of this more simplified game is similar to the full game, indicating that snakes and ladders have a complementary role in advancing and inhibiting player progress. In **Figures 8C**, simulations evaluated the removal of all ladders (except O), and a separate set of simulations evaluated the removal of all snakes (except X) in **Figure 8D**. As expected, removal of almost all ladders significantly lengthened the game such that only ~20% of games were completed in 200 spins or less. Alternatively, removal of almost all snakes shortened game duration such that 95% of games were completed in 200 spins or less.

The influence of snakes towards winning the game was determined by removing individual snakes and simulating the game for comparison with the full game as shown in **Figure 9A**. No detectable relationship was observed between the length of snakes and the number of spins that they add to a particular game on average. The most significant snakes were P, S, and T, all of which added about 15 spins to each game on average. All remaining snakes only contributed <10 additional spins per game, and snake W increased the speed of play towards victory. Snake W is in a unique location that returns players to a lower position on

___

the board such that they can thereafter land on ladders I, J, and K that take them to the top squares. In contrast, removal of ladders of the 84-square snakes and ladders game generally resulted in longer games with longer ladders. In particular, ladder A associated with knowledge and mediation had almost no impact on the length of the game, despite being 37 tiles long. Alternatively, ladder J associated with restraint and ascendance to Vijaya heaven had a disproportionate effect on slowing the progress of players; as a 38-tile ladder, it added ~56 spins on average to the length of the game when removed from the game.

*Game Board Interpretation.* All three game boards reflect the values of their creators, which reflect on the cultures and beliefs of those creators. Without the association of spiritual or behavioral elements with the snakes and ladders of the board, the game becomes a generic racing game to reach the end. But the game designers have specifically associated elements of the game with aspects of real life. All of the ladders, snakes, and chutes are associated with behaviors, penalties, and rewards as described in **Tables 1-3**. However, a challenge for interpreting the game elements is apparent in the analysis of each game board. Upon inspection, the contribution of each snake, ladder, or chute can be interpreted directly from the number of squares by which that element advances or returns a player. In other words, we interpret the importance of each game element ascribed to it by the designer of the board proportionally to its board length. Yet this proportionality between game element length and its importance only exists to a limited extent, as shown in the many of the chutes, ladders, and snakes evaluated in **Figures 3, 6, and 9**. Some of the game boards do exhibit a strong correlation between chute length and impact towards slowing progress towards victory; for example the 1998 Milton Bradley Chutes and Ladders is nearly linear in **Figure 3A and 3C**. Alternatively, the contribution of ladders to advancing players towards victory of the 72 square board in **Figure 6B** is *inversely* proportional to the length of the ladder; the longest ladder, B, actually slows progress of players. Placement of the game element also reflects importance, which when integrated into the board can be difficult to determine by inspection.

Should we interpret the intent of the game board designer and their associated cultural associations with the length of the game element; or is the actual gameplay role of an element more reflective of the game designers' intent? The game designers of the historical games did not have access to modern computational power required to simulate thousands of games, but they likely played the designed board a sufficient number of times to observe the actual contribution of each game element. But without deeper knowledge of the history of the considered boards, the game designers, and their thinking as it pertains to game board design, there will exist at least two interpretations of the significance of each game element: (i) the straightforward game element length determined by the number of tiles, and (ii) the actual contribution of each game element. Simulations conducted in this work reveal the latter, while the former is readily apparent from counting tiles.

*Game Elements of Milton Bradley 1998 Game Board.* The most important chutes by all interpretations of the modern game are P, L, and N, all of which are food related. Players (presumably children) will learn from the significant of these chutes that stealing cookies (P), eating too many chocolates (L), or running with dishes (N) that these are behaviors associated with significant penalty. In contrast, more active behaviors such as reading comic books (J), riding a bicycle without hands (O), and frolicking in the rain without a raincoat are only minor bad behaviors. With the relatively straightforward design of this board for children, the longest chutes are also the most significant for slowing player progress, making it straightforward to understand the relationship between gameplay of chutes and associated behaviors.

The ladders of the modern game are mostly insignificant, excepting ladders (I) and (E) which contribute ~20 and ~10 added spins, respectively to the game when removed. Both of these longer ladders are associated with the positive treatment of animals. Ladder E depicts a child rescuing a cat from a tree, while ladder I shows two children presenting a dog at a pet show. The importance of caring for animals is also observed in chute S, which depicts a child pulling a cat's tail and being imposed with a significant penalty of ~4 added spins to achieve victory. It is therefore surprising that ladder D, associated with caring for an injured dog, has almost no impact on the gameplay. Despite this inconsistency, children playing the game will interpret caring for animals as a positive behavior.

*72-Square Vaisnava Board.* The 72-square Vaisnava chart has varied through history such that it has been divided into five types with variations in gameplay identified by Schmidt-Madsen as type *a*, *b*, *c*, *d*,

___

and *e*, of which type *a* was identified as more significant (and considered in **Figure 4**).[27] This game board originated in western India and spread both south and north even to Nepal.[27] The history of this variation of the game, its evolution and spread throughout Asia and the world, and its spiritual significance have been extensively reviewed and discussed by Schmidt-Madsen and others.[27] The board is immediately identified as spiritual in nature by its name, Vaishnavism being one of the Hindu denominations, as well as the winning square #68 being Vaikuntha or 'Visnu's heaven.' On each side are the realms of Siva (#67) and Brahma (#69), such that all three highlighted deities represent the creation, preservation, and destruction of the world.[27] The following short description derives from more extensive description and interpretation by Schmidt-Madsen.[27] At the beginning of the game, players start on square 1 as 'birth' and proceed in the first level of Earth (square #5) and all of the tails of the snakes associated with anger (#3), greed (#4), bewilderment (#6), intoxication (#7), jealousy (#8), and desire (#9). The vertical column from Earth (square #5) up to Visnu's Heaven stand for the cosmography of Vaisnava, with the first three squares representing the Earth, atmosphere (#14), and sky/heaven (#23).Thereafter are the four higher realms including the realms of majesty (#32), men (#41), austerity (#50), and truth (#59). Hell is located in the middle of the chart in square #35, which players can occupy but immediately leave and progress through their life.

An interesting feature of the 72-square Vaisnava board is the placement of the winning square, Vaikuntha #68, before the end of the board game. For players that unfortunately move spaces beyond #68, they will continue on in their journey to either brahmlok (#69, the realm of Brahma) or the squares of satogun (#70), rajogun (#71), or tamogun (#72) corresponding to the qualities of truth, activity, and inertia, respectively. Once landing on the head of a snake on square #72, the player returns to square #51 corresponding to prithvi or 'earth,' after which they could land on square #55 (egoity) and return back down to square #2 (phenomenal reality) reliving almost the entire sequence of the game. As noted by Schmidt-Madsen, "the cyclical nature of the chart as a whole suggest that it does not just represent a journey through a single lifetime, but through an entire cycle of rebirth from entrance to exit."[27] The snakes and ladders are therefore the karmic links between periods of lives and represent transitions associated with the content of starting and ending squares.

The importance of the rebirth cycle via snake T was determined via simulation by determining the frequency of use of snake T. As shown in **Figure 10A**, only 16.8% or 12.3% of games end without players ever using snake T when rolling a six-sided die or 7 Cowrie shells, respectively; the majority of games proceed with players using snake T more than one time, making it a key element of the game. When all game elements are removed except snake T (**Figure 5B & 5F**), there exist only a small set of games that rapidly complete in less than 24 spins/rolls, while a significant number of games of more than 24 spins/rolls include the player cycling multiple times to eventually land on the winning tile #68.

The cyclical nature of the Vaisnava board with multiple cycles of players working through the board has a significant impact on the importance of ladders. Without any ladders, the game exhibits a large distribution of lengths extending out far beyond 300 spins/rolls as shown in **Figure 5C & 5G**. Most surprisingly, the effect of individual ladders towards the number of spins/rolls to achieve victory is inversely proportional to the ladder length, as shown in **Figure 6B & 6D**. Shorter ladders more significantly speed the player to victory than longer ladders, and ladder B actually slows player progress. Placement of ladders can be more important than ladder length. Despite the apparent reward of ascension of players to the Realm of Brahma in square #69 from square #17 associated with compassion, ladder B results in players taking longer to achieve victory due to their placement past the winning square of #68. The most important ladder, J, is only 14 tiles long and is twice as influential toward game progress as any other ladder; not surprisingly, ladder J associated with the virtue of devotion elevates a player from square #54 to #68 in Visnu's Heaven and instant victory.

Alternatively, as shown in **Figure 6A & 6C**, the contribution of snakes towards gameplay correlates well with snake length with a few exceptions; longer snakes add the most spins/rolls to the gameplay. The longest snakes L, M, and K are all longer than 40 tiles and all contribute on average five or more spins/rolls to the game length. These three snakes are all associated with significant vices and penalties: (i) snake L moves players behaving foolishly from square #61 to #13 associated with intermediate space, (ii) snake M moves players with egoity from square #55 to #2 associated with phenomenal reality, and (iii) the most

---

important snake K moves players associated with darkness from square #63 to #3 associated with anger. These influential snakes all move players from the upper realms back down to the lowest levels associated with the earth and the earliest stages of life. The aberration is snake N, which upon injury moves players from square #52 to #35 associated with hell; a short trip to hell benefits the player mildly and speeds their advance to victory when playing with a six-sided die (no benefit of hell was observed with 7 Cowrie shells).

General trends in the Vaisnava 72-square board exist in the importance of cycles and the weight of different classes of vices and virtues. Repeating the board after missing the chance to go to Visnu's Heaven at square #68 is important through most game elements. Knowledge is only treated as a moderately important component of life; ladder G associated with knowledge and snake O associated with ignorance only have a moderate impact on the game. Similarly, righteousness (ladder F) and unrighteousness (snake P) are only moderately important. The least important vices are all associated with relationships with others including envy (snake S), hatred (R), bad company (Q), while the most important vices (darkness, foolishness, and egoity) of snakes L, K, and M all relate to the personal feelings. By this analysis, playing through this 72-square board many times would likely lead players to value repeating their life and minimizing personal failures.

*84-Square Jaina Board.* The 84-square Jaina board depicted in **Figure 7** has a structure representative of the universe; as described by Schmidt-Madsen in greater detail,[27] the playable portion of the board depicts a human with legs, torso, arms and head consistent with a 'cosmic man' frequently depicted in Jaina art of the 16th century and later.[30] Within the torso is the grid of the main portion of the game, vertical and horizontal lines break up the body into individual units or tiles that each represent the next stage in progression.[31] The central column of the game has been identified as the planes of particular realms associated with the surrounding tiles of the row. As stated by Schmidt-Madsen, "Rows one and two represent the lower world, rows three through six represent the middle world, and rows seven through nine represent the upper world."[27] Above the torso is a six-square grid that represents the heavens, at the top of which is the winning square described as the *isatpragbhara*, or 'crescent-shaped place of perfection.' Outside the main grid on either side are tiles #56 and #66, which possibly represent the arms of the cosmic man extending across multiple realms. Finally, at the bottom below the main board is tile #1, which is associated with 'permanent basic lifeforms' that have not yet entered into the cycle of rebirth associated with the game progression.

This 84-square board form of the game represents the progression of the player out of the initial basic lifeform, which itself requires rolling a '1' on the stick, into the cycle of rebirth and growth through the different realms of the board ultimately achieving the final square of the heavens.[27] This progression begins among the 'hell-beings' of square #2 along with a row of squares representing anger, greed, delusion, pride, and jealousy. Progressing to the 2nd row brings the player among a group of ten gods that live in palaces in the 'jewel-colored hell.'[32] The third row brings the player to the level of five kinds of stationary beings, after which the 4th row is occupied by beings deficient of all of their senses. The fifth row celebrates the second rebirth of plants and beings, after which the 6th row is a third rebirth category associated with humans. The final three rows before the heavens represent the upper world and the gods who live in flying palaces. Above row #9 are the heavens, at which point players are guaranteed liberation in the final square with continual progress towards top square 6, the plane of liberation. Additional insight into the origin, labeling, and interpretation of all squares and game board elements is provided by Schmidt-Madsen.[27]

The complexity of the 84-square Jaina board compared to other versions significantly reduces the correlation between snake and ladder length and impact on speed to victory (i.e., average number of spins/rolls per game). As shown in **Figure 9A**, the longest snake, Q associated by egoity by activity, is only the fourth most important snake in terms of average added spins. The most important snakes are P, S, and T associated with five false views, black karmic stain (kṛṣṇa leśyā), and seven vices. And snake W associated with 'not practicing restraint' and the penalty of 'red karmic stain' speeds players to victory, because it returns players to the fifth level where they can retry to access ladders I, J, and K which lead to the heavens. The vices of envy (snake R) and 'not observing vows' (snake V) have almost no impact on the game, and overall, there exists no general trend in the types vices that significantly affect gameplay.

---

Regarding ladders, most have only negligible impact on the length of gameplay as shown in **Figure 9B**. Only ladders I and J associated with omniscience and purified restraint significantly advance a player forward, as these are the only two ladders the move players from the lower realms up into the heavens of the stop six squares. Of note is the long ladder A, associated with the virtue of knowledge leading to virtuous meditation, having negligible impact on the game, despite being one of the longest ladders in the game at 37 squares. The only other significant ladder is G, whereby virtuous meditation leads to omniscience. Based on the extensive number of simulations, it is apparent that players would over time interpret the importance of godly characteristics (e.g., all-knowing) versus more human characteristics of limited knowledge, contemplations, stages of purification, or discriminating judgement.

**Conclusions.** The game of Gyan Chauper or 'snakes and ladders' combines together elements of a racing style game with spiritual or behavioral concepts that communicate to the players their relative importance weighted by the design of the board elements. Without the ability to know the original intent of the game designers, the spiritual or behavior concepts associated with each snake/chute or ladder was quantified by simulation of the game boards hundreds of thousands of times with variation in the exclusion of individual elements. This analysis indicated that while many snakes and ladders contributed to the progression towards victory commensurate with the length of that game element (i.e., number of tiles), many game elements were either more or less important than can be determined by inspection. The unique importance of any game element resulted from its placement and interaction with the end game, be it a ladder directly to the winning square or a snake or ladder that resolved or ended the cyclical movement of players that overshoot the winning square. These particularly influential game elements exhibited patterns in the types of behavioral or spiritual vices and virtues they represented, possibly indicating the importance of those cultural and religious concepts to the game designer and their culture at the time of board creation.

**Acknowledgements.** The authors are grateful to the authors that came before them to find, describe, and analyze the different historical board games. We thank Jacob Schmidt-Madsen for feedback on the manuscript. We also thank Kristeen Joseph, Saurabh Maduskar, and Gaurav Kumar for valuable feedback.

**Supporting Information.** The supporting information file contains copies of Python simulation code, summary tables of data of simulation results, and instructions to access the online digital data sets.

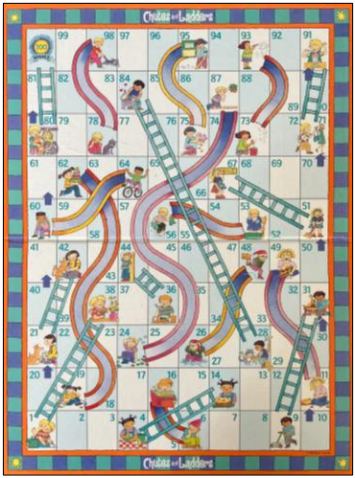 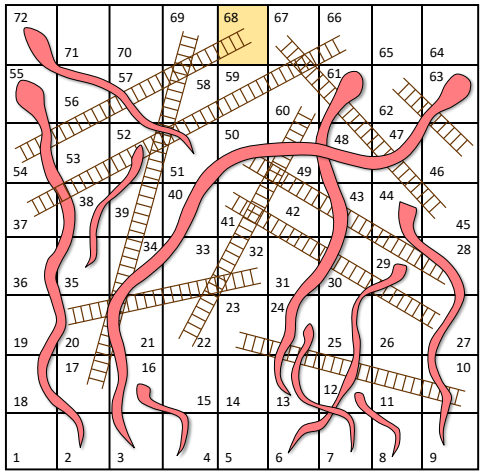 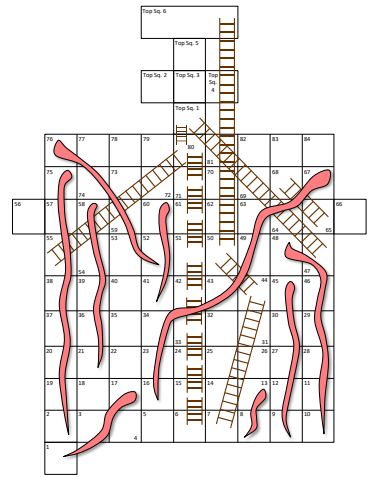

**Scheme 1.** **(A)** The 100-square Milton Bradley board game from 1998. Photograph by Paul Dauenhauer. **(B)** The 72-square Vaisnava board game. **(C)** Uninscribed 84-square Jaina chart Rajasthan, 19th century, including the top squares 1-6.

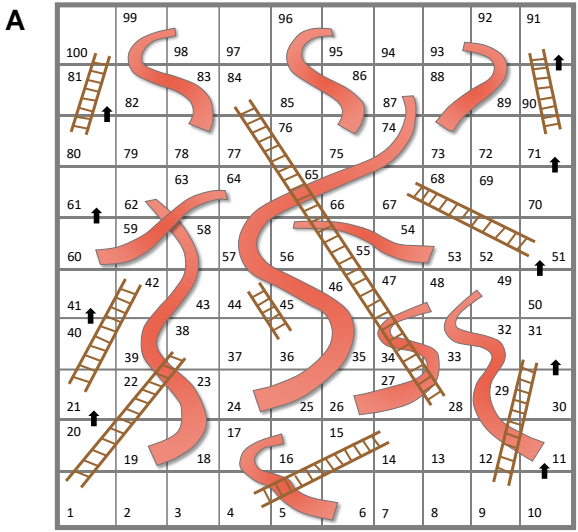
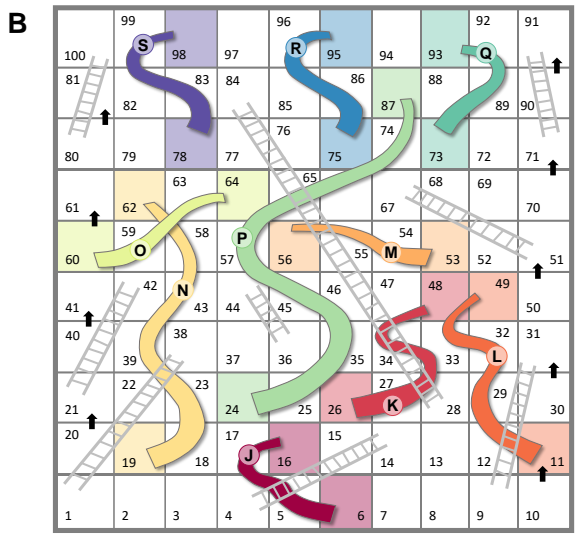
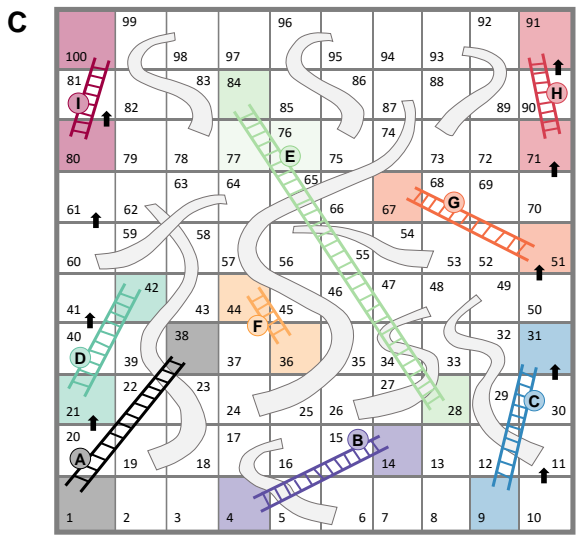

**Figure 1. Chutes and Ladders Game Board, 1998 Milton Bradley Edition.** (**A**) Gameplay squares increment in integers from 1 to 100 with ladders and snakes/chutes connecting independent squares. (**B**) Snakes/chutes J-S connect independent squares and cross through intermediate squares. (**C**) Ladders A-I connect independent squares and cross through intermediate squares.

**Table 1.** Summary of chutes and ladders and their associated virtues and vices in the 1998 Milton Bradley board 100 square game.

| Ladder | Connection | Virtue | Reward | Chute | Connection | Vice | Penalty |
|---|---|---|---|---|---|---|---|
| A | 1 to 38 | Planted a garden | Fresh flowers | J | 16 to 6 | Reading comic books | Detention |
| B | 4 to 14 | Baked a cake | Gets a cake | K | 48 to 26 | Skating on thin ice | Fall through ice |
| C | 9 to 31 | Mow the lawn | Go to the circus | L | 49 to 11 | Ate too many chocolates | Stomach ache |
| D | 21 to 42 | Care for a dog | Happy dog | M | 56 to 53 | Didn't wear coat in the rain | Became sick |
| E | 28 to 84 | Rescue a cat | Happy cat | N | 62 to 19 | Runs with dishes | Broken dishes |
| F | 36 to 44 | Ate healthy food | Growth in height | O | 64 to 60 | Bicycle riding without hands | Broken arm |
| G | 51 to 67 | Swept the floor | Go to cinema | P | 87 to 24 | Hand in the cookie jar | Injured head |
| H | 71 to 91 | Found a lost purse | Eat ice cream | Q | 93 to 73 | Drew on the walls | Wash the walls |
| I | 80 to 100 | Goes to pet show | Won a pet show | R | 95 to 75 | Broke a window | Pay for window |
|  |  |  |  | S | 98 to 78 | Pulled a cat's tail | Cat scratches |

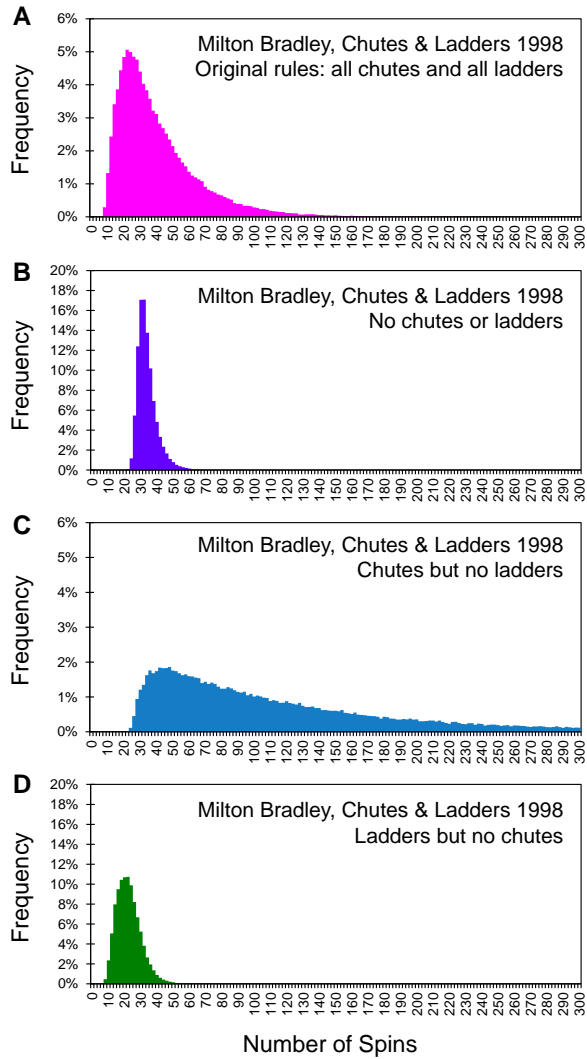

**Figure 2. Monte Carlo Probabilistic Simulation of Milton Bradley, Chutes & Ladders, 1998.** One hundred thousand iterations of the game were simulated with a single player randomly spinning a wheel with values of one-to-six proceeding to square 100. The 1998 game rules applied with: **(A)** both chutes and ladders, **(B)** without chutes or ladders, **(C)** with chutes but without ladders, and **(D)** with ladders but without chutes.

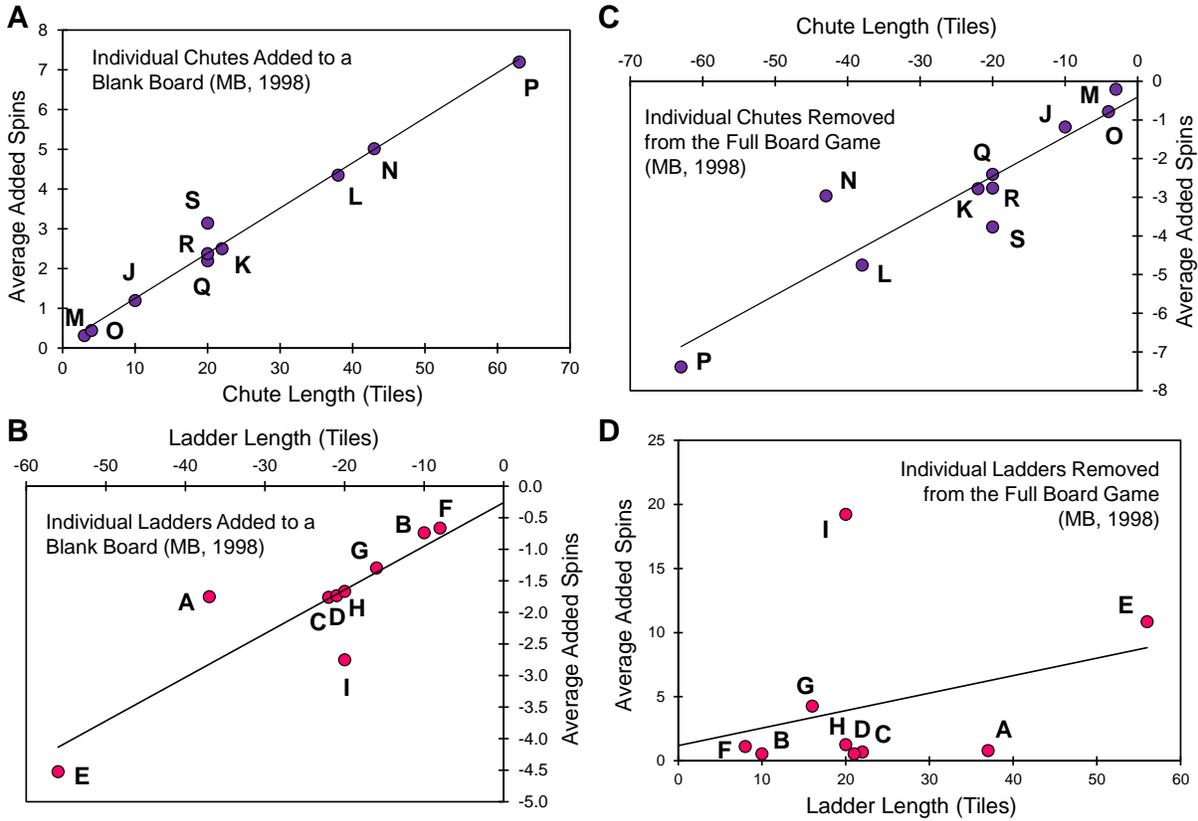

**Figure 3. Contribution of Individual Chutes and Ladders to Game Progression, Milton Bradley 1998.** Monte Carlo simulations of 100,000 games evaluated the addition of each individual chute (**A**) or ladder (**B**) to determine the number of average spins added or subtracted in comparison to a 100-square board without any chutes or ladders. Additionally, Monte Carlo simulations of 100 games evaluated the removal of each individual chute (**C**) or ladder (**D**) to determine the number of average spins subtracted or added in comparison to the full game with all chutes and all ladders.

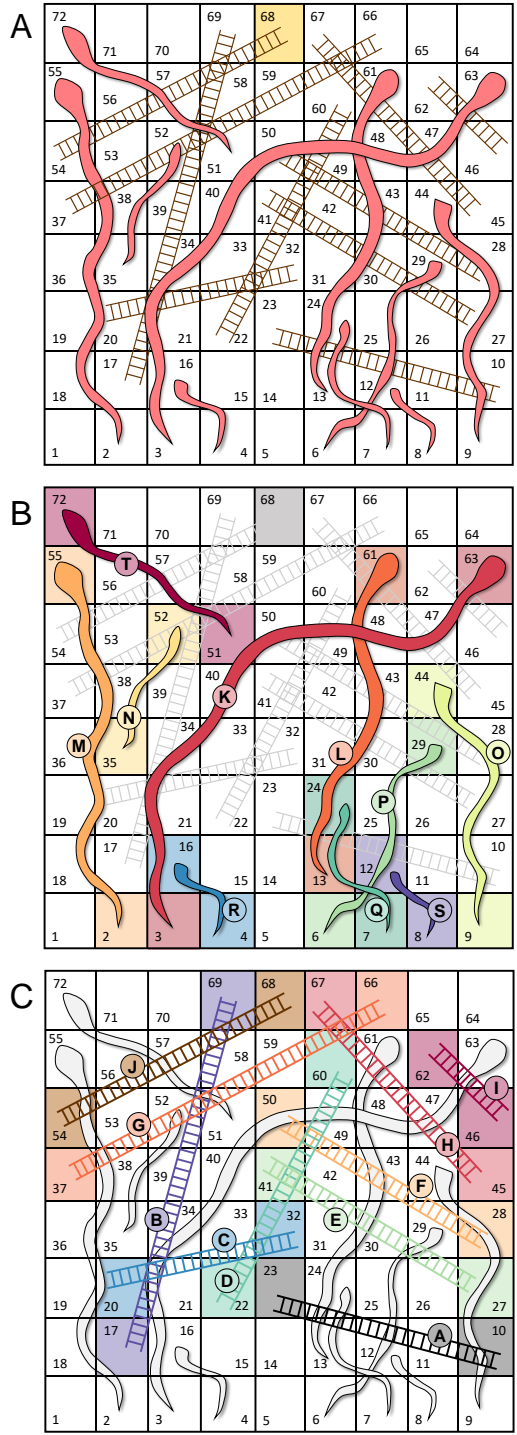

**Figure 4. Snakes and Ladders Game Board, 72-Square Edition. (A)** Gameplay squares increment in integers from 1 to 72 with ladders and snakes connecting independent squares. Players win the game by landing on square 68 highlighted in yellow. **(B)** Snakes K-T connect independent squares and cross through intermediate squares. **(C)** Ladders A-J connect independent squares and cross through intermediate squares.

**Table 2.** Summary of snakes and ladders and their associated virtues and vices in the 72 square Vaisnava game.

| Ladder | Connection | Virtue | Reward | Chute | Connection | Vice | Penalty |
|---|---|---|---|---|---|---|---|
| A | 10 to 23 | Austerity | Heaven | T | 72 to 51 | Quality of Inertia | Earth |
| B | 17 to 69 | Compassion | Realm of Brahma | K | 63 to 3 | Darkness | Anger |
| C | 20 to 32 | Charity | Realm of Majesty | L | 61 to 13 | Foolish | Intermediate Space |
| D | 22 to 60 | Religion | Intelligence | M | 55 to 2 | Egoity | Phenomenal Reality |
| E | 27 to 41 | Highest Truth | Realm of Men | N | 52 to 35 | Injury | Hell |
| F | 28 to 50 | Righteousness | Realm of Austerities | O | 44 to 9 | Ignorance | Desire |
| G | 37 to 66 | Knowledge | Bliss | P | 29 to 6 | Unrighteousness | Bewilderment |
| H | 45 to 67 | Right Knowledge | Realm of Siva | Q | 24 to 7 | Bad Company | Intoxication |
| I | 46 to 62 | Judgement | Happiness | R | 16 to 4 | Hatred | Greed |
| J | 54 to 68 | Devotion | Visnu's Heaven | S | 12 to 8 | Envy | Jealousy |

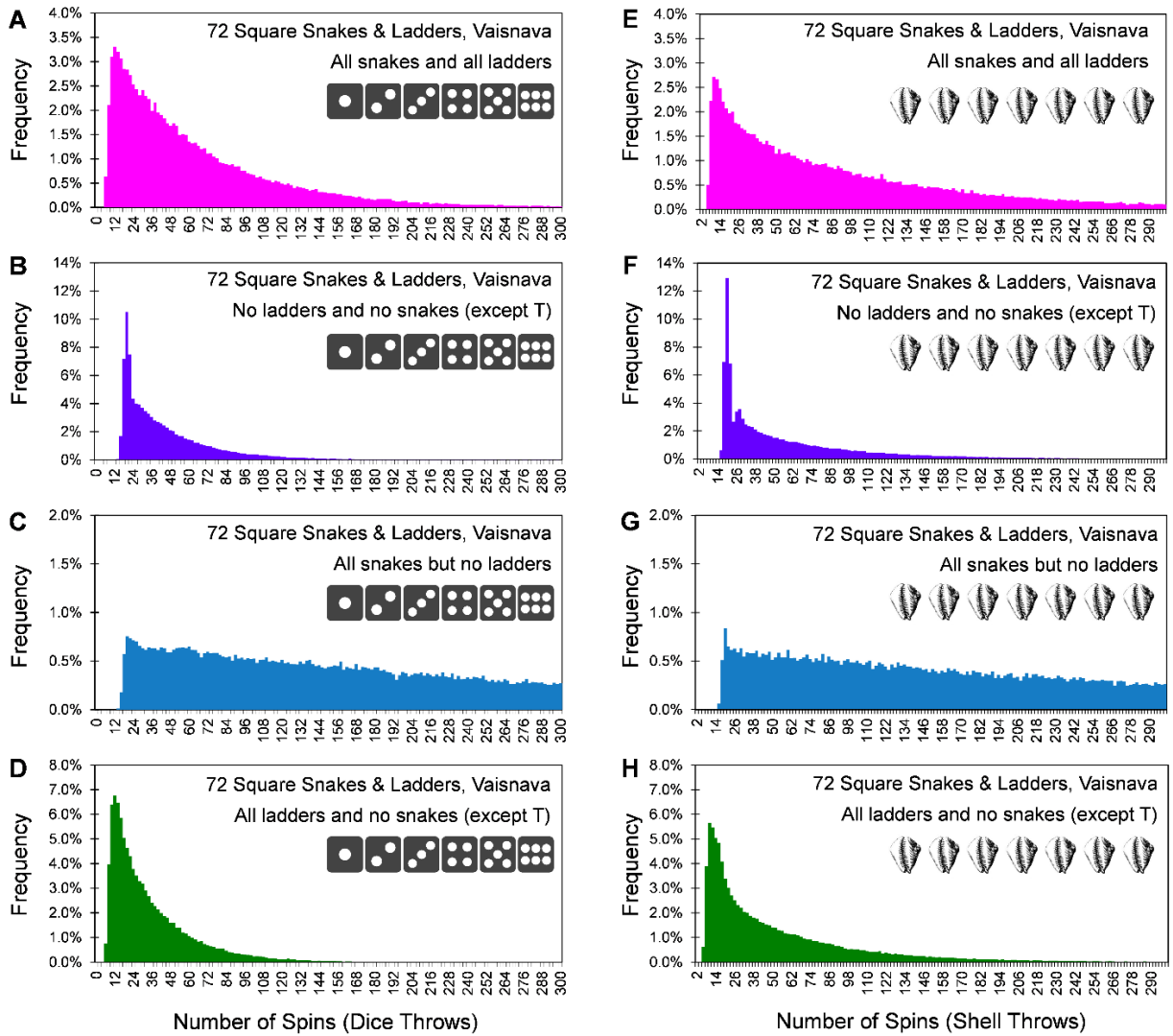

**Figure 5. Monte Carlo Probabilistic Simulation of 72-Square Snakes & Ladders using a Die or Cowrie Shells**. One hundred thousand iterations of the game were simulated with a single player randomly throwing a die (A-D, values of 1-6) or cowrie shells (E-H, values of 0-7) proceeding to square 68 for victory. The game rules applied with: **(A, E)** both snakes and ladders, **(B, F)** without ladders or snakes except T, **(C, G)** with snakes but without ladders, and **(D, H)** with ladders but without snakes except T.

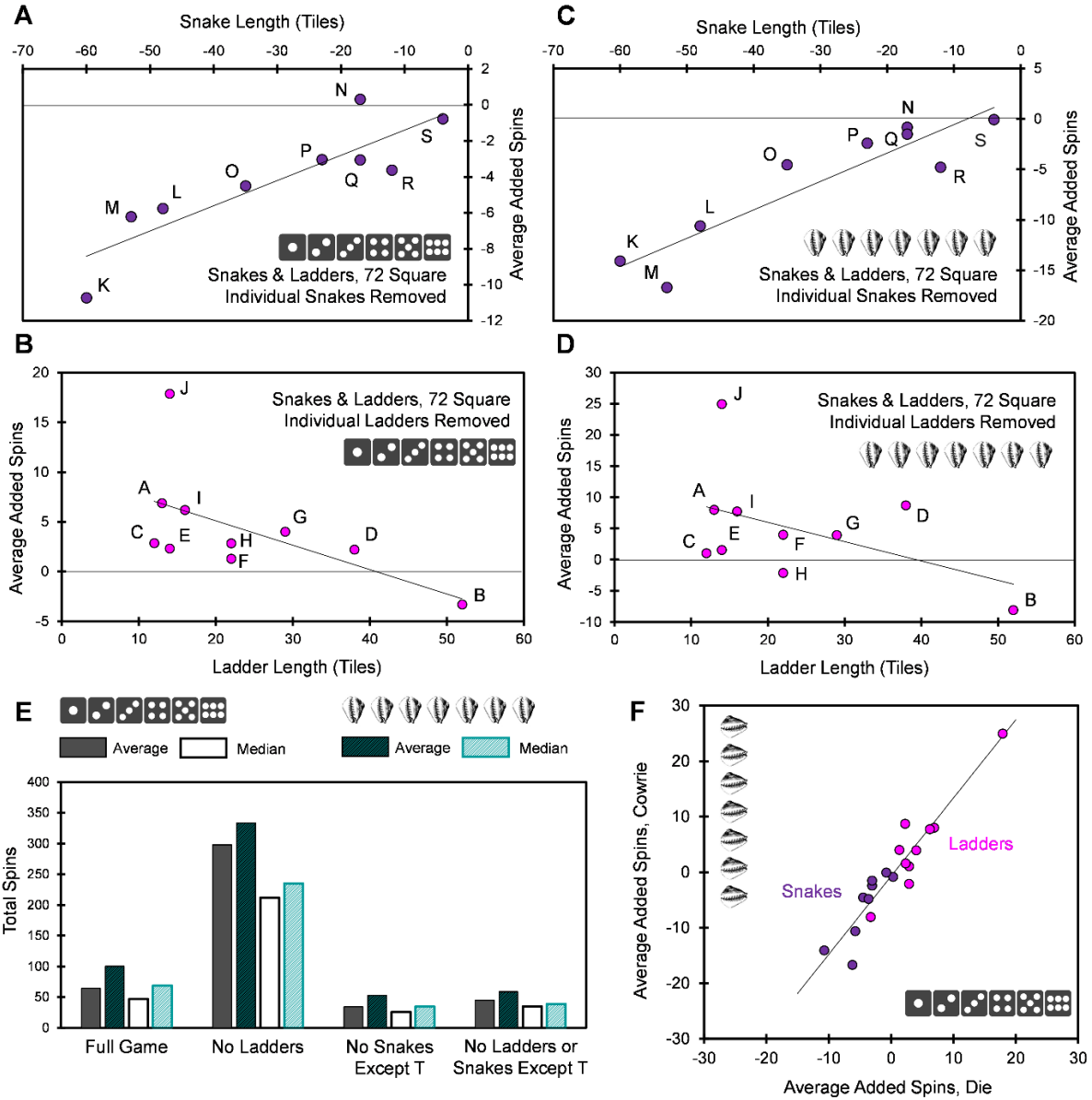

**Figure 6. Contribution of Individual Snakes and Ladders to Game Progression, 72 Square Board.** Monte Carlo simulations of 100,000 games evaluated the removal of each individual snake or ladder with a six-sided die **(A, B)** or seven Cowrie shells **(C, D)** to determine the number of average spins added in comparison to the full game with all snakes and all ladders. **(E)** Average and median total spins using a six-sided die or seven Cowrie shells for scenarios of the complete game, no ladders, no snakes except T, or no ladders or snakes except T. **(F)** Parity comparison of the average number of added spins for individually-removed ladders or snakes using seven Cowrie shells versus a six-sided die.

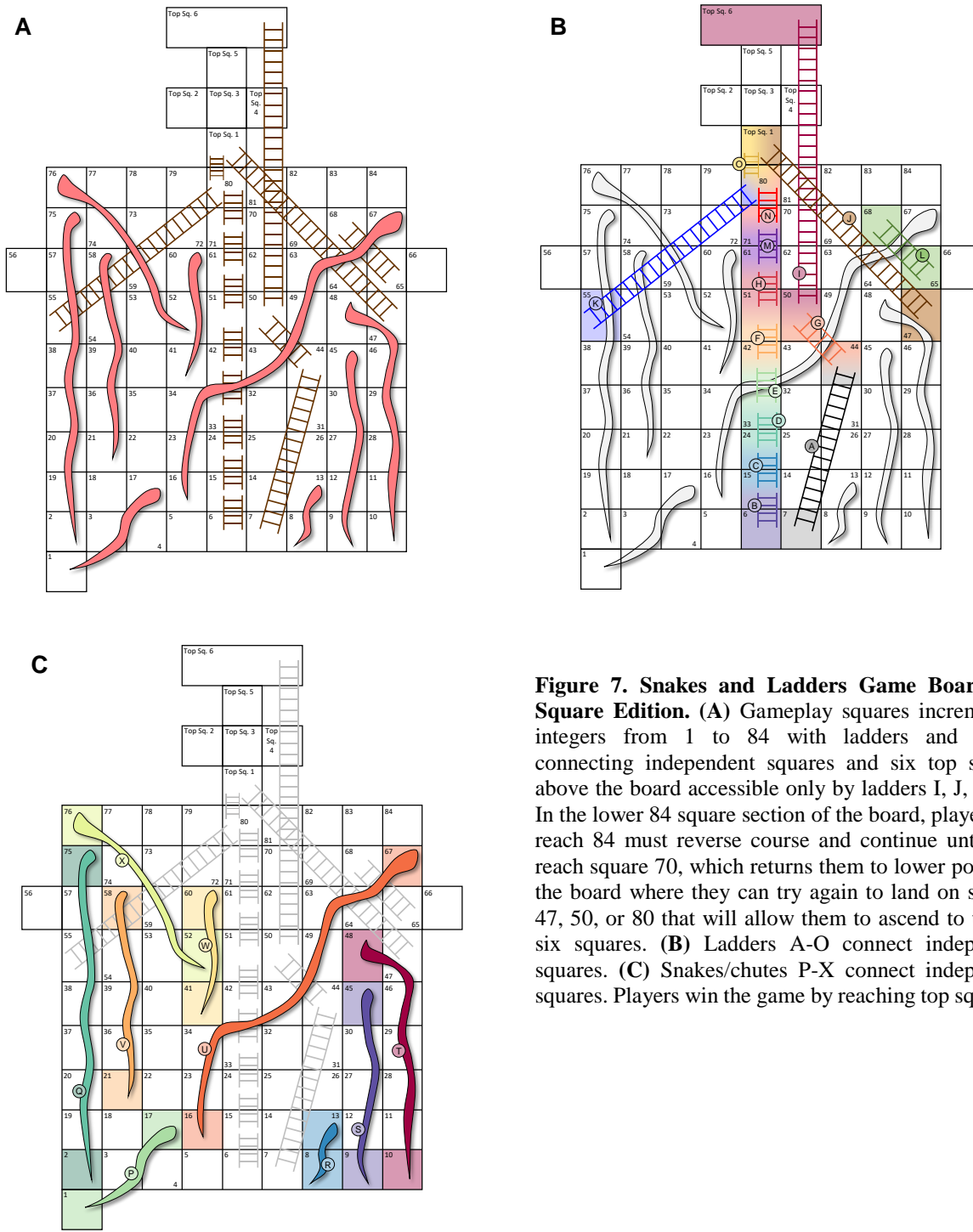

**Figure 7. Snakes and Ladders Game Board, 84-Square Edition. (A)** Gameplay squares increment in integers from 1 to 84 with ladders and snakes connecting independent squares and six top squares above the board accessible only by ladders I, J, and O. In the lower 84 square section of the board, players that reach 84 must reverse course and continue until they reach square 70, which returns them to lower points on the board where they can try again to land on squares 47, 50, or 80 that will allow them to ascend to the top six squares. **(B)** Ladders A-O connect independent squares. **(C)** Snakes/chutes P-X connect independent squares. Players win the game by reaching top square 6.

**Table 3.** Summary of snakes/chutes and ladders and their associated virtues and vices in the 84-square game board.

| Ladder | Connection | Virtue | Reward | Snake | Connection | Vice | Penalty |
|---|---|---|---|---|---|---|---|
| A | 7 to 44 | knowledge | virtuous mediattion | P | 17 to 1 | five false views | infinitely existing basic lifeforms |
| B | 6 to 15 | purification stages 1-3 | purification stages 4-6 | Q | 75 to 2 | egoity by activity | forms of hell-beings |
| C | 15 to 24 | purification stages 4-6 | purification stages 7-9 | R | 13 to 8 | envious | jelousy |
| D | 24 to 33 | purification stages 7-9 | purification stages 10-12 | S | 45 to 9 | black karmic stain | pride |
| E | 33 to 42 | purification stages 10-12 | purification stage 13 | T | 48 to 10 | seven vices | deceit from ignorance |
| F | 42 to 51 | purification stage 13 | purification stage 14 | U | 67 to 16 | egoity by inertia | dvipa- & udadhi-kumara gods |
| G | 44 to 50 | virtuous meditation | omniscience | V | 58 to 21 | not observing vows | non-permanent basic lifeforms |
| H | 51 to 61 | purification stage 14 | heavens, 400,000 gods | W | 60 to 41 | not practicing restraint | red karmic stain |
| I | 50 to TS6 | omniscience | crystalline plane of liberation | X | 76 to 52 | deluding karma | injuring another |
| J | 47 to TS1 | purified restraint | Vijaya heaven | | | | |
| K | 55 to 80 | 12 contemplations, 10 proper conducts | Sumanans & Priyadarsana neck heavens | | | | |
| L | 65 to 68 | discriminating judgement | one year of desired attainments | | | | |
| M | 61 to 71 | heavens, 400,000 gods | heavens & souls | | | | |
| N | 71 to 80 | heavens, capable and incapable souls | Sumanans & Priyadarsana neck heavens | | | | |
| O | 80 to TS1 | Sumanans & Priyadarsana neck heavens | TS1: Vijaya heaven | | | | |

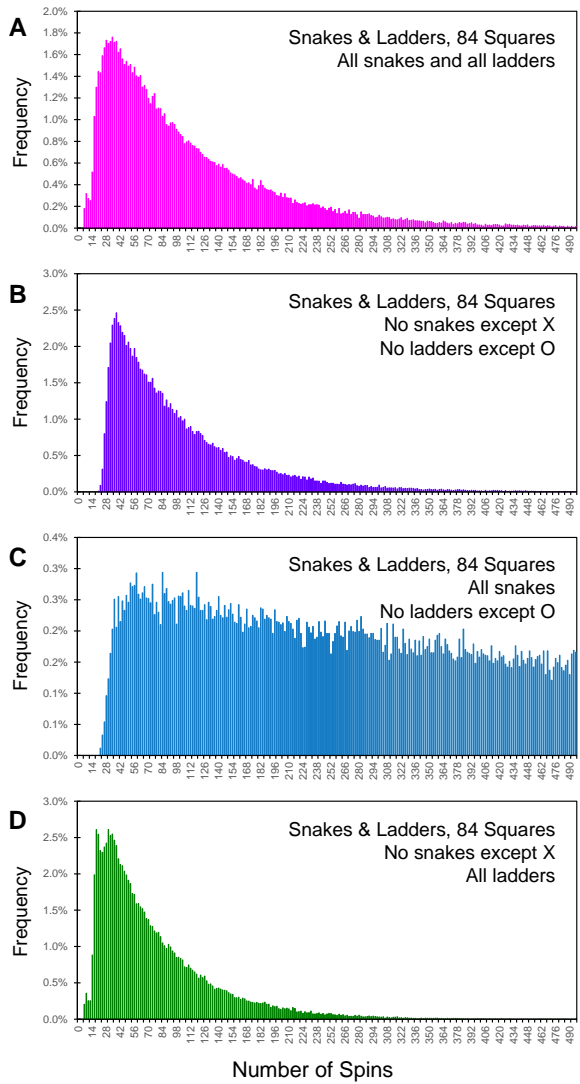

**Figure 8. Monte Carlo Probabilistic Simulation of 84-Square Snakes & Ladders.** One hundred thousand iterations of the game were simulated with a single player randomly throwing a die or cowrie shells with values of 1, 2, 5 or 6 proceeding to square TS6 for victory. The game rules applied with: **(A)** both snakes and ladders, **(B)** without ladders except O or snakes except X, **(C)** with snakes but without ladders except O, and **(D)** with all ladders but without snakes except X.

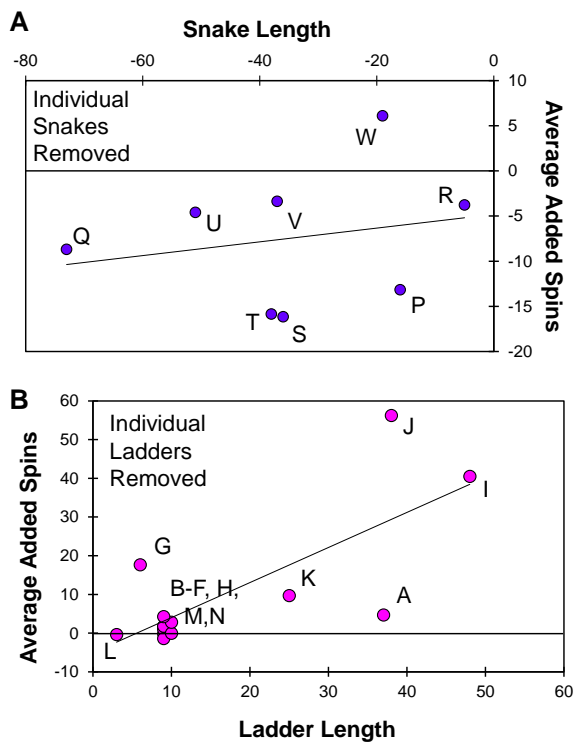

**Figure 9. Contribution of Individual Snakes and Ladders to Game Progression, 84 Square Board.** Monte Carlo simulations of 100,000 games evaluated the removal of each individual snake **(A)** or ladder **(B)** to determine the number of average spins added in comparison to the full game with all snakes and all ladders. Ladder O and snake X were retained in all simulations. The removal of a single ladder lengthened the game requiring additional spins for almost all simulations, while removal of a single snake shortened the game requiring few additional spins for all simulations except snake W.

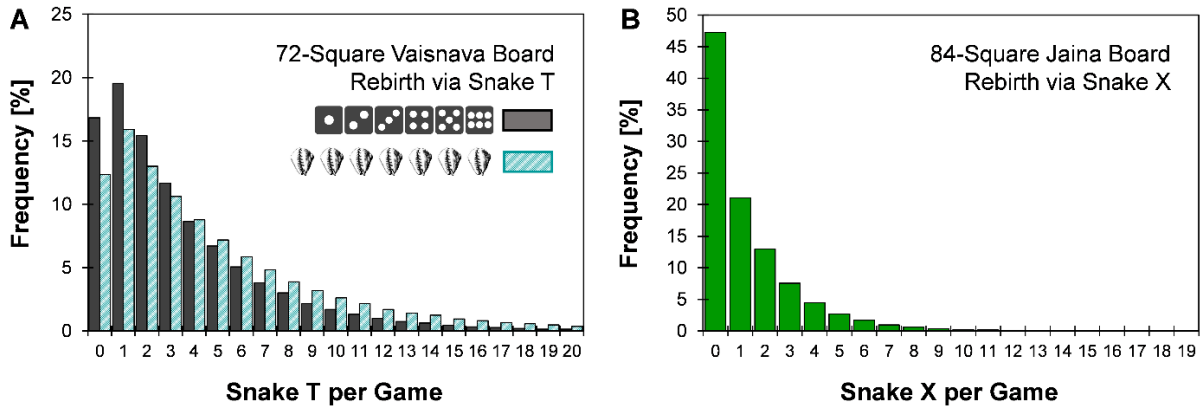

**Figure 10. Frequency of rebirth in the Vaisnava and Jaina board games. (A)** The frequency of players using snake T in the 72-square Vaisnava board using either a six-sided die (black) or seven Cowrie shells (teal stripes) indicates that the majority of games have the player go through a rebirth via snake T multiple times per game. **(B)** The frequency of players using snake X in the 84-square Jaina board indicates that 47.2% of players never use snake X, but a small majority of players do use it one or more times.

# SUPPORTING INFORMATION
# Quantifying the Vices and Virtues of Snakes and Ladders Through Time

Eleanor A. Dauenhauer[1], Paul J. Dauenhauer[2,*]

[1] Chippewa Middle School, 5000 Hodgson Road Connection, St. Paul, MN, USA 55126
[2] Department of Chemical Engineering & Materials Science, University of Minnesota, 421 Washington Ave. SE, Minneapolis, MN, USA 55455
[*] Corresponding author: hauer@umn.edu

**Table of Contents**

# 1. Game Simulation Python Code, 1998 Chutes & Ladders

```python
#Chutes and Ladders Game
import random
import array as arr   # importing "array" for array creations
import numpy as np
import csv
totalnumberofgames = 100000  # sets the number of total games that will be played
numberofgamesprogress = 0  #sets the total number of gampes played so far to zero
nameofgamefile = "BaseGame_100000.txt"     #determine the name of the file to
which we are writing data

f = open(nameofgamefile, 'w')
#f.write("Base Chutes and Ladders Game")
f.close()

while (numberofgamesprogress < totalnumberofgames):     #start of all games

    #Player starts off the board at position zero
    player1position = 0
    endofgame = 0
    spinvalue = 0
    numberofspins = 0

    while (endofgame<1):              # start of one game
        numberofspins = numberofspins + 1
        spinvalue = random.randint(1,6)
        if spinvalue + player1position < 101:
            player1position = player1position + spinvalue

        #check to see if player won the game
        if player1position == 100:
            endofgame = 10

        #ladders
        if player1position == 1:        #Ladder A
            player1position = 38

        if player1position == 4:        #Ladder B
            player1position = 14

        if player1position == 9:        #Ladder C
            player1position = 31

        if player1position == 21:       #Ladder D
```

```python
        player1position = 42

    if player1position == 28:       #Ladder E
        player1position = 84

    if player1position == 36:       #Ladder F
        player1position = 44

    if player1position == 51:       #Ladder G
        player1position = 67

    if player1position == 71:       #Ladder H
        player1position = 91

    if player1position == 80:       #Ladder I
        player1position = 100

    #chutes
    if player1position == 16:       #Chute J
        player1position = 6

    if player1position == 48:       #Chute K
        player1position = 26

    if player1position == 49:       #Chute L
        player1position = 11

    if player1position == 56:       #Chute M
        player1position = 53

    if player1position == 62:       #Chute N
        player1position = 19

    if player1position == 64:       #Chute O
        player1position = 60

    if player1position == 87:       #Chute P
        player1position = 24

    if player1position == 93:       #Chute Q
        player1position = 73

    if player1position == 95:       #Chute R
        player1position = 75
```

```python
        if player1position == 98:          #Chute S
            player1position = 78

    #print("Position:", player1position)         #Prints out number of spins in a game
    #print("Number of spins:", numberofspins)
    converted_numberofspins = str(numberofspins)

    # Opening a file and saving data
    f = open(nameofgamefile, 'a')
    f.writelines(converted_numberofspins)
    #f.writelines("|")
    f.writelines("\n")    #make it go to next line
    f.close()

    numberofgamesprogress = numberofgamesprogress + 1    #increment the number of games played

print("done")
```

## 2. Summary data sets, 1998 Chutes & Ladders

**Table S1.** Monte Carlo simulation of 1998 Milton Bradley Chutes and Ladders with the full game, removal of all chutes and ladders, removal of all ladders, and removal of all chutes. Numbers are the number of spins (1-6) required to reach square 100. Total of 100,000 game trials.

|          | Full Game | No Chutes / No Ladders | No Ladders | No Chutes |
|----------|-----------|------------------------|------------|-----------|
| Average  | 40.1      | 33.3                   | 120.2      | 22.0      |
| St. Dev. | 25.6      | 6.1                    | 89.9       | 7.6       |
| Max      | 318       | 93                     | 1041       | 84        |
| Q3       | 51        | 36                     | 154        | 26        |
| Median   | 33        | 32                     | 93         | 21        |
| Q2       | 22        | 29                     | 56         | 16        |
| Min      | 7         | 20                     | 21         | 7         |

**Table S2.** Monte Carlo simulation of 1998 Milton Bradley Chutes and Ladders with the full game compared with individual removal of ladders from the full game. Numbers are the number of spins (1-6) required to reach square 100. Total of 100,000 game trials. Delta tiles is the number of tiles connecting the beginning and end of the game element.

|            | Full Game | Remove A | Remove B | Remove C | Remove D | Remove E | Remove F | Remove G | Remove H | Remove I |
|------------|-----------|----------|----------|----------|----------|----------|----------|----------|----------|----------|
| Average    | 40.1      | 40.9     | 40.6     | 40.8     | 40.7     | 51.0     | 41.2     | 44.4     | 41.4     | 59.3     |
| Delta Avg  |           | **0.8**  | **0.5**  | **0.7**  | **0.6**  | **10.9** | **1.1**  | **4.3**  | **1.3**  | **19.2** |
| Delta Tiles|           | **37**   | **10**   | **22**   | **21**   | **56**   | **8**    | **16**   | **20**   | **20**   |
| St. Dev.   | 25.6      | 25.6     | 25.6     | 25.9     | 26.0     | 34.7     | 26.4     | 28.6     | 26.6     | 42.1     |
| Max        | 318       | 318      | 294      | 350      | 340      | 392      | 297      | 373      | 328      | 446      |
| Q3         | 51        | 52       | 51       | 51       | 51       | 65       | 52       | 56       | 52       | 76       |
| Median     | 33        | 34       | 34       | 34       | 34       | 41       | 34       | 37       | 34       | 47       |
| Q2         | 22        | 23       | 23       | 23       | 23       | 27       | 23       | 24       | 23       | 30       |
| Min        | 7         | 7        | 7        | 7        | 7        | 7        | 7        | 7        | 7        | 7        |

**Table S3.** Monte Carlo simulation of 1998 Milton Bradley Chutes and Ladders with the full game compared with individual removal of chutes from the full game. Numbers are the number of spins (1-6) required to reach square 100. Total of 100,000 game trials. Delta tiles is the number of tiles connecting the beginning and end of the game element.

|  | Full Game | Remove J | Remove K | Remove L | Remove M | Remove N | Remove O | Remove P | Remove Q | Remove R | Remove S |
|---|---|---|---|---|---|---|---|---|---|---|---|
| Average | 40.1 | 38.9 | 37.3 | 35.4 | 39.9 | 37.1 | 39.3 | 32.7 | 37.7 | 37.3 | 36.3 |
| Delta Avg |  | **-1.2** | **-2.8** | **-4.8** | **-0.2** | **-3.0** | **-0.8** | **-7.4** | **-2.4** | **-2.8** | **-3.8** |
| Delta Tiles |  | **-10** | **-22** | **-38** | **-3** | **-43** | **-4** | **-63** | **-20** | **-20** | **-20** |
| St. Dev. | 25.6 | 24.9 | 23.1 | 21.1 | 25.5 | 22.4 | 24.8 | 16.3 | 23.1 | 22.8 | 21.6 |
| Max | 318 | 258 | 315 | 224 | 345 | 292 | 275 | 159 | 257 | 288 | 262 |
| Q3 | 51 | 49 | 47 | 44 | 50 | 46 | 50 | 41 | 47 | 47 | 45 |
| Median | 33 | 32 | 31 | 30 | 33 | 31 | 33 | 29 | 32 | 31 | 31 |
| Q2 | 22 | 22 | 21 | 21 | 22 | 22 | 22 | 21 | 22 | 21 | 21 |
| Min | 7 | 7 | 7 | 7 | 7 | 7 | 7 | 7 | 7 | 7 | 7 |

## 3. Game simulation Python code, 72-square Vaisnava board, Die

```python
#Chutes and Ladders Game
import random
import array as arr   # importing "array" for array creations
import numpy as np
import csv
totalnumberofgames = 100000  # sets the number of total games that will be played
numberofgamesprogress = 0  #sets the total number of gampes played so far to zero
reborn = 0    #sets the total number of times player is reborn at zero
nameofgamefile = "72_square_100000.txt"    #determine the name of the file to
which we are writing data

f = open(nameofgamefile, 'w')
#f.write("Base Chutes and Ladders Game")
f.close()

while (numberofgamesprogress < totalnumberofgames):     #start of all games

    #Player starts off the board at position zero
    player1position = 0
    endofgame = 0
    spinvalue = 0
    numberofspins = 0
    reborn = 0

    while (endofgame<1):            # start of one game
        numberofspins = numberofspins + 1
        spinvalue = random.randint(1,6)
        if spinvalue + player1position < 73:
            player1position = player1position + spinvalue

        #check to see if player won the game
        if player1position == 68:
            endofgame = 10

        #ladders
        if player1position == 10:        #Ladder A
            player1position = 23

        if player1position == 17:        #Ladder B
            player1position = 69

        if player1position == 20:        #Ladder C
            player1position = 32
```

```python
        if player1position == 22:        #Ladder D
            player1position = 60

        if player1position == 27:        #Ladder E
            player1position = 41

        if player1position == 28:        #Ladder F
            player1position = 50

        if player1position == 37:        #Ladder G
            player1position = 66

        if player1position == 45:        #Ladder H
            player1position = 67

        if player1position == 46:        #Ladder I
            player1position = 62

        if player1position == 54:        #Ladder J
            player1position = 68

        #chutes
        if player1position == 72:        #Chute T, REQUIRED
            player1position = 51
            reborn = reborn + 1          #counts the number of times players go down the chute at 72

        if player1position == 63:        #Chute K
            player1position = 3

        if player1position == 61:        #Chute L
            player1position = 13

        if player1position == 55:        #Chute M
            player1position = 2

        if player1position == 52:        #Chute N
            player1position = 35

        if player1position == 44:        #Chute O
            player1position = 9

        if player1position == 29:        #Chute P
            player1position = 6
```

```python
        if player1position == 24:        #Chute Q
            player1position = 7

        if player1position == 16:        #Chute R
            player1position = 4

        if player1position == 12:        #Chute S
            player1position = 8

    #print("Position:", player1position)        #Prints out number of spins in a game
    #print("Number of spins:", numberofspins)
    converted_reborn = str(reborn)

    # Opening a file and saving data
    f = open(nameofgamefile, 'a')
    f.writelines(converted_reborn)
    #f.writelines("|")
    f.writelines("\n")      #make it go to next line
    f.close()

    numberofgamesprogress = numberofgamesprogress + 1    #increment the number of games played

print("done")
```

## 3.1. Game simulation Python code, 72-square Vaisnava board, Cowrie Shells with Determination of Rebirth by Snake/Chute T

```python
#Chutes and Ladders Game
import random
import array as arr   # importing "array" for array creations
import numpy as np
import csv
totalnumberofgames = 100000  # sets the number of total games that will be played
numberofgamesprogress = 0  #sets the total number of gampes played so far to zero
nameofgamefile = "72_square_cowrie_Resurrection.txt"     #determine the name of
the file to which we are writing data

f = open(nameofgamefile, 'w')
#f.write("Base Chutes and Ladders Game")
f.close()

while (numberofgamesprogress < totalnumberofgames):    #start of all games

    #Player starts off the board at position zero
    player1position = 0
    endofgame = 0
    spinvalue = 0
    numberofspins = 0
    rebirth = 0

    while (endofgame<1):              # start of one game
        numberofspins = numberofspins + 1
        spinvalue = random.randint(0,1) + random.randint(0,1) +
random.randint(0,1) + random.randint(0,1) +random.randint(0,1) +
random.randint(0,1) + random.randint(0,1)        #cowrie shells allow for 0 to
7
        #print(spinvalue)
        if spinvalue + player1position < 73:
            player1position = player1position + spinvalue

        #check to see if player won the game
        if player1position == 68:
            endofgame = 10

        #ladders
        if player1position == 10:        #Ladder A
            player1position = 23

        if player1position == 17:        #Ladder B
            player1position = 69
```

```python
    if player1position == 20:        #Ladder C
        player1position = 32

    if player1position == 22:        #Ladder D
        player1position = 60

    if player1position == 27:        #Ladder E
        player1position = 41

    if player1position == 28:        #Ladder F
        player1position = 50

    if player1position == 37:        #Ladder G
        player1position = 66

    if player1position == 45:        #Ladder H
        player1position = 67

    if player1position == 46:        #Ladder I
        player1position = 62

    if player1position == 54:        #Ladder J
        player1position = 68

#chutes
    if player1position == 72:        #Chute T, REQUIRED
        rebirth = rebirth + 1        #increments rebirth
        player1position = 51

    if player1position == 63:        #Chute K
        player1position = 3

    if player1position == 61:        #Chute L
        player1position = 13

    if player1position == 55:        #Chute M
        player1position = 2

    if player1position == 52:        #Chute N
        player1position = 35

    if player1position == 44:        #Chute O
        player1position = 9
```

```python
        if player1position == 29:       #Chute P
            player1position = 6

        if player1position == 24:       #Chute Q
            player1position = 7

        if player1position == 16:       #Chute R
            player1position = 4

        if player1position == 12:       #Chute S
            player1position = 8

    #print("Position:", player1position)        #Prints out number of spins in a game
    #print("Number of spins:", numberofspins)
    converted_rebirth = str(rebirth)

    # Opening a file and saving data
    f = open(nameofgamefile, 'a')
    f.writelines(converted_rebirth)
    #f.writelines("|")
    f.writelines("\n")     #make it go to next line
    f.close()

    numberofgamesprogress = numberofgamesprogress + 1    #increment the number of games played

print("done")
```

## 3.2 Cowrie Shell Statistics

The throwing of Cowrie shells (also spelled 'Cowry') allows for a distribution of scores for each throw associated with the number of shells that land up or down. By throwing seven Cowrie shells, there exist 8 possible throwing scores associated with all shells up, all shells down, and all combinations in between for a total score options of X = 0, 1, 2, 3, 4, 5, 6, or 7.  The simulation used a probability of 50% (p = 0.50) and a total of seven shells (n = 7), which leads to binomial probability distribution described by equation S1.

$$P(X) = \frac{n!}{(n-X)!X!} \, p^X \, (1-p)^{(n-X)} \qquad (Eq.\ S1)$$

**Table S4.** Bionomial probability distribution for the throwing of 7 Cowrie shells with 50%/50% probability (p) of landing up or down.

| Cowry Value, X | Binomial Distribution |
| --- | --- |
| 0 | 0.78% |
| 1 | 5.47% |
| 2 | 16.41% |
| 3 | 27.34% |
| 4 | 27.34% |
| 5 | 16.41% |
| 6 | 5.47% |
| 7 | 0.78% |

The assumption of a 50% probability for Cowrie shells landing up or down was based on prior work experimentally evaluating the throwing of four Cowrie shells. As shown in Table S5, the probability of throwing four Cowrie shells for total values of 0, 1, 2, 3, and 4 were measured experimentally as 6.6%, 24.3%, 38.1%, 23.6%, and 7.4%, respectively. This data was fit to a binomial distribution according to equation S1 to minimize the absolute value of the difference with a value of p of 50%.

**Table S5.** Experimental probability data for throwing four Cowrie shells fit to a binomial distribution (p = 0.50).  Source:

| Cowry Value, X | Probability Data | Binomial Distribution | Difference |
| --- | --- | --- | --- |
| 0 | 0.066 | 0.0625 | 0.0035 |
| 1 | 0.243 | 0.25 | 0.007 |
| 2 | 0.381 | 0.375 | 0.006 |
| 3 | 0.236 | 0.25 | 0.014 |
| 4 | 0.074 | 0.0625 | 0.0115 |

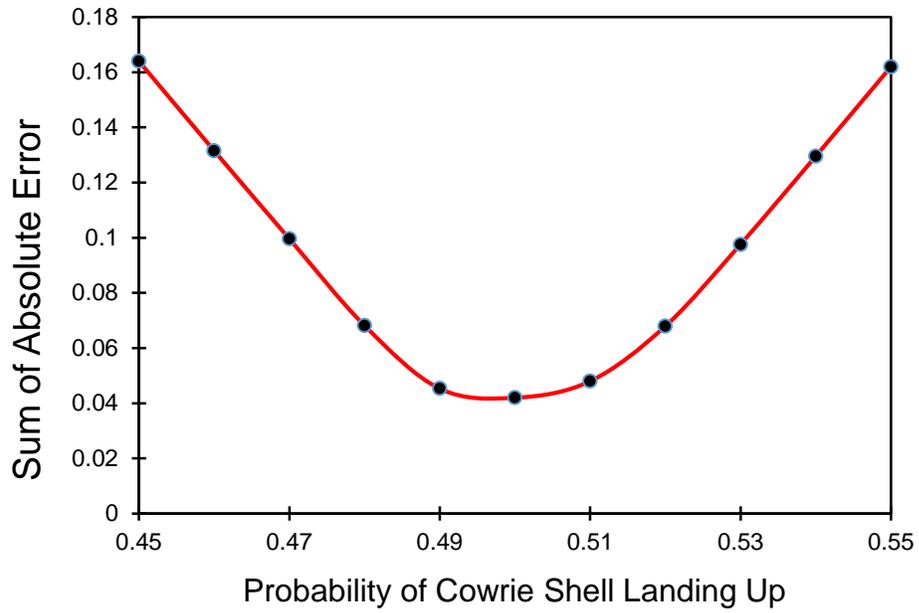

**Figure S1.** Sum of absolute error of binomial distribution model versus experimental data with variable probability.

# 4. Summary data sets, 72-square Vaisnava board

**Table S4.1.** Monte Carlo simulation of 72-Square Vaisnava board using a die (values of 1-6) with the full game, removal of all chutes and ladders, removal of all ladders, and removal of all chutes. Snake T was always left in all simulated game boards to allow for players to proceed after overshooting the winning square. Numbers are the number of spins (1-6) required to reach square 68. Total of 100,000 game trials.

|  | Full 72 Game | No Ladders | No Ladders or Snakes except T | No Snakes except T |
|---|---|---|---|---|
| Average | 64.5 | 297.9 | 44.9 | 34.6 |
| St. Dev. | 57.6 | 282.0 | 31.2 | 28.0 |
| Max | 648 | 3176 | 358 | 326 |
| Median | 47 | 212 | 35 | 26 |
| Min | 5 | 14 | 13 | 5 |

**Table S5.1.** Monte Carlo simulation of 72-Square Vaisnava board using a die (values of 1-6) with the full game compared with individual removal of **ladders** from the full game. Numbers are the number of spins (1-6) required to reach square 68. Total of 100,000 game trials. Delta tiles is the number of tiles connecting the beginning and end of the game element.

|  | Full Game | Remove A | Remove B | Remove C | Remove D | Remove E | Remove F | Remove G | Remove H | Remove I | Remove J |
|---|---|---|---|---|---|---|---|---|---|---|---|
| Average | 64.5 | 71.4 | 61.2 | 67.4 | 66.8 | 66.9 | 65.8 | 68.6 | 67.4 | 70.7 | 82.4 |
| St. Dev. | 57.6 | 63.1 | 54.7 | 60.6 | 60.0 | 59.7 | 58.9 | 61.4 | 60.3 | 63.5 | 76.5 |
| Delta tiles |  | 13.0 | 52.0 | 12.0 | 38.0 | 14.0 | 22.0 | 29.0 | 22.0 | 16.0 | 14.0 |
| Delta Avg. |  | 6.9 | -3.3 | 2.9 | 2.2 | 2.3 | 1.3 | 4.0 | 2.9 | 6.2 | 17.9 |
| Max | 648 | 758 | 642 | 883 | 717 | 669 | 684 | 784 | 673 | 969 | 1230 |
| Median | 47 | 52 | 45 | 49 | 48 | 48 | 48 | 50 | 49 | 51 | 59 |
| Min | 5 | 6 | 5 | 5 | 5 | 5 | 5 | 5 | 5 | 5 | 5 |

**Table S6.1.** Monte Carlo simulation of 72-Square Vaisnava board using a die (values of 1-6) with the full game compared with individual removal of **snakes** from the full game. Numbers are the number of spins (1-6) required to reach square 68. Total of 100,000 game trials. Delta tiles is the number of tiles connecting the beginning and end of the game element. Snake T was always left in all simulated game boards to allow for players to proceed after overshooting the winning square.

|  | Full Game | Remove K | Remove L | Remove M | Remove N | Remove O | Remove P | Remove Q | Remove R | Remove S |
|---|---|---|---|---|---|---|---|---|---|---|
| Average | 64.5 | 53.8 | 58.8 | 58.3 | 64.8 | 60.0 | 61.5 | 61.5 | 60.9 | 63.7 |
| St. Dev. | 57.6 | 46.4 | 51.6 | 50.9 | 58.5 | 53.5 | 55.3 | 54.8 | 54.6 | 57.1 |
| Delta tiles |  | -60.0 | -48.0 | -53.0 | -17.0 | -35.0 | -23.0 | -17.0 | -12.0 | -4.0 |
| Delta Avg. |  | -10.7 | -5.8 | -6.2 | 0.3 | -4.5 | -3.0 | -3.1 | -3.6 | -0.8 |
| Max | 648 | 544 | 514 | 643 | 775 | 795 | 754 | 756 | 660 | 675 |
| Median | 47 | 40 | 43 | 43 | 47 | 44 | 44 | 45 | 44 | 46 |
| Min | 5 | 5 | 5 | 5 | 5 | 5 | 5 | 5 | 5 | 5 |

**Table S4.2.** Monte Carlo simulation of 72-Square Vaisnava board using 7 Cowrie shells (values of 0-7) with the full game, removal of all chutes and ladders, removal of all ladders, and removal of all chutes. Snake T was always left in all simulated game boards to allow for players to proceed after overshooting the winning square. Tabulated values are the number of spins (0-7) required to reach square 68. Total of 100,000 game trials.

|  | Full 72 Game | No Ladders | No Snakes except T | No Ladders or Snakes except T |
|---|---|---|---|---|
| Average | 100.5 | 333.4 | 52.9 | 59.0 |
| Standard Deviation | 99.1 | 318.2 | 51.8 | 52.2 |
| Minimum | 5 | 15 | 5 | 14 |
| Median | 69 | 235 | 35 | 39 |
| Maximum | 1326 | 4025 | 532 | 577 |

**Table S5.2.** Monte Carlo simulation of 72-Square Vaisnava board using 7 Cowrie shells (values of 0-7) with the full game compared with individual removal of **ladders** from the full game. Numbers are the number of spins (0-7) required to reach square 68. Total of 100,000 game trials. Delta tiles is the number of tiles connecting the beginning and end of the game element.

|  | Full Game | Remove A | Remove B | Remove C | Remove D | Remove E | Remove F | Remove G | Remove H | Remove I | Remove J |
|---|---|---|---|---|---|---|---|---|---|---|---|
| Average | 100.5 | 108.5 | 92.4 | 101.5 | 109.2 | 102.0 | 104.5 | 104.4 | 98.4 | 108.2 | 125.4 |
| St. Dev. | 99.1 | 105.1 | 91.2 | 99.8 | 106.6 | 99.0 | 103.0 | 102.1 | 96.4 | 105.7 | 123.0 |
| Delta Tiles |  | 13 | 52 | 12 | 38 | 14 | 22 | 29 | 22 | 16 | 14 |
| Delta Avg. |  | 8.0 | -8.1 | 1.1 | 8.7 | 1.6 | 4.0 | 4.0 | -2.1 | 7.8 | 25.0 |
| Min | 5 | 6 | 5 | 5 | 5 | 5 | 5 | 5 | 5 | 5 | 5 |
| Median | 69 | 76 | 63 | 70 | 76 | 71 | 72 | 72 | 68 | 75 | 87 |
| Max | 1326 | 1145 | 1236 | 1056 | 1275 | 1044 | 1307 | 1136 | 1265 | 1300 | 1567 |

**Table S6.2.** Monte Carlo simulation of 72-Square Vaisnava board using 7 Cowrie shells (values of 0-7) with the full game compared with individual removal of **snakes** from the full game. Numbers are the number of spins (0-7) required to reach square 68. Total of 100,000 game trials. Delta tiles is the number of tiles connecting the beginning and end of the game element. Snake T was always left in all simulated game boards to allow for players to proceed after overshooting the winning square.

|  | Full Game | Remove K | Remove L | Remove M | Remove N | Remove O | Remove P | Remove Q | Remove R | Remove S |
|---|---|---|---|---|---|---|---|---|---|---|
| Average | 100.5 | 86.4 | 89.9 | 83.8 | 99.6 | 95.9 | 98.0 | 98.9 | 95.7 | 100.4 |
| St. Dev. | 99.1 | 84.4 | 87.8 | 81.5 | 97.9 | 94.6 | 96.8 | 97.5 | 94.4 | 98.9 |
| Delta Tiles |  | -60 | -48 | -53 | -17 | -35 | -23 | -17 | -12 | -4 |
| Delta Avg. |  | -14.1 | -10.6 | -16.7 | -0.8 | -4.5 | -2.4 | -1.5 | -4.8 | -0.1 |
| Min | 5 | 5 | 5 | 5 | 5 | 5 | 5 | 5 | 5 | 5 |
| Median | 69 | 59 | 62 | 58 | 68 | 66 | 67 | 68 | 65 | 69 |
| Max | 1326 | 1063 | 1028 | 1051 | 1141 | 1151 | 1114 | 1115 | 1180 | 1060 |

## 5. Game simulation Python code, 84-square Jaina board

```python
#Chutes and Ladders Game
import random
import array as arr    # importing "array" for array creations
import numpy as np
import csv
totalnumberofgames = 100000   # sets the number of total games that will be played
numberofgamesprogress = 0   #sets the total number of gampes played so far to zero
nameofgamefile = "84_square_100000.txt"      #determine the name of the file to which we are writing data

f = open(nameofgamefile, 'w')
#f.write("Base Chutes and Ladders Game")
f.close()

while (numberofgamesprogress < totalnumberofgames):      #start of all games, loops for each game

    #Player starts off the board at position one
    player1position = 1                         #is the current square of the player 1-84. Six top squares 93-98. Simulated reverse squares 85-92
    endoflowergame = 0                          #tracks if the lower game has been won 1-92
    endofhighergame = 0                         #tracks if higher game has been won 93-98
    spinvalue = 0                               #is the value of the current spin
    numberofspins = 0                           #tracks the number of spins in a game
    toploop = 0

    while (endoflowergame<1):                   # start of lower game for spaces 1 to 92

        endofturn = 0                           #tracks if the turn is over.  0 means turn still in progress.  1 means turn is over

        #starting position cycles. Only occurs on position 1.
        if player1position == 1:                #If the player is on the starting position
            while(player1position < 2):         #Continue looping until the player gets off square 1
                numberofspins = numberofspins + 1    #Counts the number of spins in the game
                spinvalue = random.randint(1,4)     #Roll 1, 2, 3, or 4
                if spinvalue == 1:              #if the player rolls a 1, then move to position 2
```

```python
                player1position = 2

        numberofspins = numberofspins + 1   #counts the number of spins in the game
        spinvalue = random.randint(1,4) #game only permits rolls of 1, 2, 5, 6.  Rolls of 3 & 4 are not possible
        if spinvalue == 3:              #converts a 3 spin to a 5 spin
            spinvalue = 5
        if spinvalue == 4:              #converts a 4 spin to a 6 spin
            spinvalue = 6

        #ladders.  If you are at the base of a ladder and rolled a 1 then you advance up the ladder
        #Ladder I needs to come before G which needs to come before A to prevent multiple ladders per turn
        if player1position == 50 and spinvalue == 1:        #Ladder I
            player1position = 98
            endofturn = 1

        if player1position == 44 and spinvalue == 1:        #Ladder G
            player1position = 50
            endofturn = 1

        if player1position == 7 and spinvalue == 1:         #Ladder A
            player1position = 44
            endofturn = 1

        #Ladder O needs to come before N and K
        if player1position == 80 and spinvalue == 1:        #Ladder O   ALWAYS KEEP ON
            player1position = 93
            endofturn = 1

        if player1position == 88 and spinvalue == 1:        #Ladder OO. This is ladder 0 if the player is going backwards ALWAYS KEEP ON
            player1position = 93
            endofturn = 1

        if player1position == 71 and spinvalue == 1:        #Ladder N
            player1position = 80
            endofturn = 1

        if player1position == 55 and spinvalue == 1:        #Ladder K
            player1position = 80
            endofturn = 1
```

```python
        #The central ladder system requires backwards order of O then N, M, H, F, E, D, C, B

        if player1position == 61 and spinvalue == 1:        #Ladder M
            player1position = 71
            endofturn = 1

        if player1position == 51 and spinvalue == 1:        #Ladder H
            player1position = 61
            endofturn = 1

        if player1position == 42 and spinvalue == 1:        #Ladder F
            player1position = 51
            endofturn = 1

        if player1position == 33 and spinvalue == 1:        #Ladder E
            player1position = 42
            endofturn = 1

        if player1position == 24 and spinvalue == 1:        #Ladder D
            player1position = 33
            endofturn = 1

        if player1position == 15 and spinvalue == 1:        #Ladder C
            player1position = 24
            endofturn = 1

        if player1position == 6 and spinvalue == 1:         #Ladder B
            player1position = 15
            endofturn = 1

        if player1position == 65 and spinvalue == 1:        #Ladder L
            player1position = 68
            endofturn = 1

        if player1position == 47 and spinvalue == 1:        #Ladder J
            player1position = 93
            endofturn = 1

        #check to see if player won the lower part of game and has ascended to 93-98
        if player1position > 92:      #top square of lower game is 92
            endoflowergame = 10
```

```python
        if spinvalue + player1position < 93 and endofturn == 0:
            player1position = player1position + spinvalue        #Advances player spinvalue number of spaces

            #chutes/snakes
            if player1position == 17:       #Chute P
                player1position = 1

            if player1position == 75:       #Chute Q
                player1position = 2

            if player1position == 13:       #Chute R
                player1position = 8

            if player1position == 45:       #Chute S
                player1position = 9

            if player1position == 48:       #Chute T
                player1position = 10

            if player1position == 67:       #Chute U
                player1position = 16

            if player1position == 58:       #Chute V
                player1position = 21

            if player1position == 60:       #Chute W
                player1position = 41

            if player1position == 76:       #Chute X ALWAYS KEEP ON
                player1position = 52

            if player1position == 92:       #Chute XX ALWAYS KEEP ON
                player1position = 52

        #print("Spin value:", spinvalue)                    #Print spin value
        #print("Player position:", player1position)         #Print out the current player1position on this turn
        #print("Number of spins:", numberofspins)           #Print out the current number of spins so far

    while (endofhighergame < 1):                 #Starts the loop for progress on squares 93-98
        numberofspins = numberofspins + 1    #Counts the number of spins in the game
```

```python
        spinvalue = random.randint(1,4)      #Roll 1, 2, 3, or 4
        if spinvalue == 3:                   #converts a 3 spin to a 5 spin
            spinvalue = 5
        if spinvalue == 4:                   #converts a 4 spin to a 6 spin
            spinvalue = 6
        if spinvalue + player1position < 99:
            player1position = player1position + spinvalue       #Advances player spinvalue number of spaces
        if player1position == 98:
            endofhighergame = 10

    #print("Position:", player1position)      #Prints out number of spins in a game
    #print("Number of spins:", numberofspins)
    converted_numberofspins = str(numberofspins)

    # Opening a file and saving data
    f = open(nameofgamefile, 'a')
    f.writelines(converted_numberofspins)
    #f.writelines("|")
    f.writelines("\n")    #make it go to next line
    f.close()

    numberofgamesprogress = numberofgamesprogress + 1    #increment the number of games played

print("done")
```

# 6. Summary data sets, 84-square Jaina board

**Table S7.** Monte Carlo simulation of 84-square Jaina board with the full game, removal of all chutes and ladders, removal of all ladders, and removal of all chutes. Ladder O and snake X was always left in all simulated game boards to allow for players to proceed after overshooting the winning square. Numbers are the number of spins (1, 2, 5, 6) required to reach top square 6. Total of 100,000 game trials.

|          | Full Game | No Ladders except O | No Ladders or Snakes except O & X | No Snakes except X |
|----------|-----------|---------------------|-----------------------------------|--------------------|
| Average  | 110.2     | 775.3               | 100.7                             | 76.4               |
| St. Dev. | 93.3      | 744.9               | 72.5                              | 61.0               |
| Max      | 1147      | 10156               | 858                               | 769                |
| Median   | 82        | 546                 | 79                                | 58                 |
| Min      | 6         | 21                  | 19                                | 6                  |

**Table S8.** Monte Carlo simulation of 84-Square Jaina board with the full game compared with individual removal of **ladders** from the full game. Numbers are the number of spins (1, 2, 5, 6) required to reach top square 6. Total of 100,000 game trials. Delta tiles is the number of tiles connecting the beginning and end of the game element. Ladder O and snake X was always left in all simulated game boards to allow for players to proceed after overshooting the winning square.

|              | Full Game | Remove ladders |       |       |       |       |       |       |       |       |       |       |       |       |       |
|              |           | A     | B     | C     | D     | E     | F     | G     | H     | I     | J     | K     | L     | M     | N     |
|--------------|-----------|-------|-------|-------|-------|-------|-------|-------|-------|-------|-------|-------|-------|-------|-------|
| Average      | 110.2     | 114.9 | 110.0 | 111.9 | 111.2 | 112.0 | 108.8 | 127.9 | 110.2 | 150.8 | 166.5 | 119.9 | 109.9 | 113.0 | 114.5 |
| St. Dev.     | 93.3      | 96.1  | 92.9  | 94.5  | 94.3  | 95.2  | 92.4  | 109.5 | 93.2  | 130.5 | 149.6 | 103.5 | 92.7  | 96.0  | 98.2  |
| Max          | 1147      | 1493  | 1151  | 1254  | 1119  | 1226  | 1212  | 1331  | 1104  | 1528  | 1722  | 1071  | 1069  | 1566  | 1255  |
| Median       | 82        | 86    | 82    | 83    | 83    | 83    | 81    | 95    | 82    | 111   | 121   | 89    | 82    | 84    | 85    |
| Min          | 6         | 11    | 6     | 6     | 6     | 6     | 6     | 6     | 6     | 7     | 6     | 6     | 6     | 6     | 6     |
| Delta Avg    |           | 4.7   | -0.2  | 1.7   | 1     | 1.8   | -1.4  | 17.7  | 0     | 40.5  | 56.2  | 9.7   | -0.4  | 2.8   | 4.3   |
| Delta Squares|           | 37    | 9     | 9     | 9     | 9     | 9     | 6     | 10    | 48    | 38    | 25    | 3     | 10    | 9     |

**Table S9.** Monte Carlo simulation of 84-Square Jaina board with the full game compared with individual removal of **snakes** from the full game. Numbers are the number of spins (1, 2, 5, 6) required to reach top square 6. Total of 100,000 game trials. Delta tiles is the number of tiles connecting the beginning and end of the game element. Ladder O and snake X was always left in all simulated game boards to allow for players to proceed after overshooting the winning square.

|  | Full Game | Snake P | Snake Q | Snake R | Snake S | Snake T | Snake U | Snake V | Snake W |
|---|---|---|---|---|---|---|---|---|---|
| Average | 110.2 | 97.1 | 101.5 | 106.5 | 94.1 | 94.4 | 105.6 | 106.9 | 116.3 |
| St. Dev. | 93.3 | 80.5 | 83.1 | 90.3 | 77.5 | 77.1 | 88.0 | 89.5 | 99.7 |
| Max | 1147 | 1011 | 931 | 1079 | 934 | 897 | 1090 | 1137 | 1199 |
| Median | 82 | 73 | 77 | 79 | 71 | 71 | 79 | 80 | 86 |
| Min | 6 | 6 | 6 | 6 | 6 | 6 | 6 | 6 | 6 |
| Delta Avg |  | -13.1 | -8.7 | -3.8 | -16.1 | -15.8 | -4.6 | -3.4 | 6.1 |
| Delta Squares |  | -16 | -73 | -5 | -36 | -38 | -51 | -37 | -19 |

# 7. Data Digital Availability in the University of Minnesota Data Repository

The raw data generated via simulation have been stored in the Data Repository of the University of Minnesota (DRUM) and are available online.

Data Set Link:
- https://conservancy.umn.edu/handle/11299/254844
- https://doi.org/10.13020/yvv6-cx09

Data Set DOI:   10.13020/yvv6-cx09

DRUM Website:   https://conservancy.umn.edu/drum